\documentclass[journal,draftcls,onecolumn,12pt,twoside]{IEEEtranTCOM}
\usepackage{cite}
\usepackage{color}
\usepackage{amsfonts,amssymb}
\usepackage{bm}
\usepackage{amsmath}
\usepackage{multirow}
\usepackage{algorithmic}
\usepackage{graphicx}
\usepackage{float}
\usepackage{subcaption}
\usepackage{textcomp}
\usepackage{graphicx} 
\usepackage{epstopdf}
\usepackage{array}
\usepackage[thmmarks,amsmath]{ntheorem}
\newtheorem{mythm}{Theorem}
\newtheorem{mydef}{Definition}
\newtheorem{assumption}{Assumption}
\newtheorem{lemma}{Lemma}

\newtheorem{corollary}{Corollary}

\usepackage{array}
\usepackage{enumerate}
\usepackage{booktabs}
\usepackage{algorithm}
\usepackage{algorithmic}
\usepackage{setspace}
\usepackage{float}
\usepackage{cases}

\qedsymbol{\ensuremath{\square}} 
{   
	\theoremstyle{nonumberplain}
	\theoremheaderfont{\bfseries}
	\theorembodyfont{\normalfont}
	\theoremsymbol{\ensuremath{\square}} 
	
}

\normalsize

\ifCLASSINFOpdf
\else
\fi

\begin{document}
%
\title{
	Coincidence analysis of Stackelberg  and Nash equilibria in three-player leader-follower security games
}
%
%
%


\author{Gehui Xu, Guanpu Chen, Zhaoyang Cheng,\\  Yiguang Hong,  and Hongsheng Qi
	\thanks{This work was supported by the National Natural Science Foundation of
		China (No. 62173250, No. 61873262), by Shanghai Municipal Science and Technology Major
		Project (No. 2021SHZDZX0100). Corresponding author:  Guanpu Chen.
	}
	\thanks{Gehui Xu, Zhaoyang Cheng, and Hongsheng Qi are with Key Laboratory of Systems and Control, Academy of Mathematics and Systems Science, Beijing, China,
		and are also with School of Mathematical Sciences, University of Chinese Academy of Sciences, Beijing, China. (e-mail: xghapple@amss.ac.cn,  chengzhaoyang@amss.ac.cn,  qihongsh@amss.ac.cn).}
	\thanks{Guanpu Chen is with JD Explore Academy,  Beijing, China. (e-mail: chengp@amss.ac.cn).}
	\thanks{Yiguang Hong is with
		Department of Control Science and Engineering, Shanghai Research Institute for Intelligent Autonomous Systems, Tongji University, Shanghai, and is also with the Key Laboratory of Systems and Control, Academy of Mathematics and Systems Science, Chinese Academy of Sciences, Beijing, China. (e-mail: yghong@iss.ac.cn).}
	
}

%
%

\markboth{IEEE Transactions on Wireless Communications}%
{Submitted paper}
%



\maketitle

\begin{abstract}
	
There has been significant recent interest in  leader-follower security games, where the leader dominates the decision process with the Stackelberg equilibrium (SE) strategy. However, such a leader-follower scheme may become invalid in practice {due to subjective or objective factors}, 
and then
 the Nash equilibrium (NE) strategy may be an alternative option. 
In this case, the leader may face a  dilemma of choosing an SE strategy or an NE strategy. In this paper, we focus on a unified three-player leader-follower security game and study the coincidence
 between SE and NE. We first explore a necessary and sufficient condition {for the case that each SE is an  NE}, which 
 can be further presented concisely when the SE is unique. This condition not only provides access  to 
   seek a satisfactory SE strategy but also makes a criterion to verify an obtained SE strategy.
 { Then we  provide another appropriate condition for the case that at least one SE is an NE.}
 Moreover, since the coincidence condition may not always be satisfied, we
 describe the closeness between SE and NE, and give an upper bound of their deviation.
   Finally, we show the applicability of the obtained theoretical results in several practical security cases, including the secure transmission problem and the cybersecurity defense.
\end{abstract}


%
\IEEEpeerreviewmaketitle

\section{Introduction}
%
%

Security games, which usually describe situations that the protected system defends against malicious attacks, have been widely applied in many fields such as  secure wireless communications, cyber-physical systems (CPS), and unmanned aerial vehicles (UAV).
The three-player security game, as one of the important categories, models the interactive details about defense or attack operations by focusing on three different types of players, with a broad range in many important security scenarios. For instance,
\cite{fang2018three} investigated 
a physical layer security issue 
among a transmitter, a relay, and an eavesdropper, and
\cite{liu2021flipit} studied an advanced persistent threat (APT) problem 
among a defender, an insider, and an attacker, while  \cite{bhattacharya2011spatial} considered a 
vehicle formation problem among  two
vehicles and a jammer.

One classical  game model to reflect players’ strategic behaviors  in  security games is 
based on leader-follower models \cite{yang2013coping,wang2015pricing,wu2018secure,garnaev2020jamming}. {
	In the models, the leader  dominates the decision process and adopts its optimal strategy by taking account into the followers' reaction, while the follower chooses the best response (BR) strategy after observing the leader’s strategy. The corresponding equilibrium  is the well-known    Stackelberg equilibrium (SE) \cite{bacsar1998dynamic}.
} In the three-player leader-follower game,  there  is  a tri-level hierarchical structure:  the top level, the middle level, and the  bottom  level.
Accordingly,
{players at high levels are called leaders, 
	while players at low levels are called followers.} 
For example, 
the  source-destination pair  at the bottom level is a follower and decides the required
transmit power  based on the observed strategies of the power station and the jammer \cite{li2019three}. Besides,   the defender at the top level is a leader and chooses its defense rate with the consideration of the attacker and the insiders’  strategies \cite{feng2015stealthy}.

However, such a  leader-follower scheme may  become invalid in practice, because  the low-level player may
lose the ability or interest to adopt the BR strategy and even ruin the leader-follower scheme for different reasons, including
the limitation of the  observation ability, the disturbance of the environment, and the stealthy of the player’s existence.  In fact,   
the jammer may have observation errors
due to
the uncertainty of the time-variant channel states  \cite{xiao2015anti};
the terrorists may choose to directly act
in consideration of the expensive surveillance cost of the defense strategy \cite{an2013security}; and
the attacker may turn to  the stealthy attack
scheme instead of the  leader-follower scheme to avoid the defender's fault detection  \cite{xiao2018dynamic}.

Hence, when the low-level player does not
strictly comply with the leader-follower scheme,
{ the high-level player will lose its corresponding dominant position, since its SE strategy is no longer the optimal one against
	the  low-level player's non-BR strategy.
}
In this view, a simultaneous-move game model may be  another acceptable description,  and the  best-known solution concept therein is the Nash equilibrium (NE), where players choose their optimal strategies independently without  observation and dominance \cite{nash1951non,korzhyk2011stackelberg,xu2022efficient}.
Since no one can benefit from changing its  strategy unilaterally, it is acceptable for the high-level player to accomplish such an NE  when its SE is not available.  In some  practical security problems, 
the high-level player may take the NE strategy   when the low-level player 
has the observation barrier \cite{altman2011jamming}, and may
tolerate an NE
to avoid an unsatisfactory outcome \cite{wu2011anti}.



 {
	Given  the above consideration, 
	a high-level player may have to face  a  dilemma: which strategy should be adopted, an SE in the  leader-follower scheme or an NE in the simultaneous-move scheme?}
Clearly, the conflict among players’ strategies under different schemes may result in the failure to achieve either SE or NE and may bring a loss in the utility for the high-level player.  
However, provided that  SE coincide with  NE,
the high-level player will not  suffer from these misgivings anymore. If so, 
the high-level player can take an SE strategy since its utility is as the same as that of taking an NE strategy. 
Moreover, when the coincidence relationship  is not satisfied,  the high-level player can still be fairly reassured of an SE strategy if the SE is quite close to an NE, and the brought gap in the high-level player’s utility 
is small and tolerated. {
	 Such analogous discussions on the relationship between SE and NE have already been a hot topic in security games,
	   and have been analyzed on two-player models such as the
	 radio transmission problem \cite{xiao2015anti} and the security deployment issue \cite{korzhyk2011stackelberg}.
}

Therefore, this paper focuses on how to help high-level players get rid of the dilemma about the strategy selection in a  three-player  security game. Specifically, we  explore the coincidence condition  when an SE is an NE. Moreover, if an effort fails, then we study
 the deviation between an SE and an NE.

\textbf{Contribution:}

{We consider a  three-player game model established for typical security  problems, including  secrecy capacity optimization \cite{wu2018secure}, cooperative secure communication \cite{fang2018three}, and APT   defense \cite{feng2015stealthy}.
Compared with existing literatures, this is the first work that studies the coincidence relationship between SE and NE under a three-player game-theoretical problem.
{Firstly, we   explore a necessary and sufficient condition such that each SE is an NE, and present its concise form when the SE is unique.}
This coincidence analysis not only 
develops an approach to seek an SE  that  exactly meets  an NE,
but also  provides a criterion  to verify whether an obtained SE is an NE.
{When sometimes not all SE are NE, we  further focus on 
 whether there exists an SE that coincides with an NE and  provide a  condition to find that at least one SE is an NE,}
  in which
the high-level players can
accurately adopt
a satisfactory 
SE strategy.
{Secondly, considering that 
the coincide relationship may not exist in all the practical situations,}
we give an upper bound of the deviation between  SE and  NE  to measure their closeness, in order to reassure the high-level player for still adopting an acceptable SE strategy. 
Finally,  we show the applicability of the obtained theoretical results in several practical security cases, including the secure transmission and the cybersecurity defense.}

%

\textbf{Related work:}

Of particular relevance to this work is the research  on three-player security games.
Accordingly, wireless communication is one of the most important fields to investigate  three-player models.
In  \cite{wu2018secure}, 
the macro base station (MBS)  employed the jamming SBSs to jam the external eavesdropping for secure transmission, while the jamming SBSs required offloading service from the helping SBSs to satisfy the users. In   \cite{li2019three}, 
the source-destination pair at the top level priced the energy transmitted to the middle-level jammer for
maximizing the secrecy rate, and the jammer decided the required transmit power to the bottom-level power station for broadcasting energy. Moreover, in \cite{fang2018three},
the source defended against the eavesdropper with the help of the relay for secure communication by  employing a leader-follower scheme. Also, there are  other fields involving three-player games.
As for  CPS security \cite{feng2015stealthy,liu2021flipit,chen2022defense},  a defender-insider-attacker game model was widely used   to study  stealthy behaviors and insider threats. In UAV  formation \cite{bhattacharya2011spatial},  
a  zero-sum game with  two vehicles and a jammer was proposed  to analyze
 mobile intruder jamming.

Another highly relevant topic to this study is 
about relationships  between SE and NE,
which has been investigated in some two-player leader-follower security games
\cite{korzhyk2011stackelberg,sengupta2019general,xiao2015anti,cheng2022single}. For instance, \cite{korzhyk2011stackelberg} considered a  security deployment issue and derived a sufficient condition related to the defender's strategic allocation  subset
such that the defender's SE strategy is also an NE strategy. Afterward, \cite{sengupta2019general} extended this condition into
a Markov game under the moving target defense background to analyze the optimal strategy for resource placement.
Moreover, \cite{xiao2015anti} compared the effectiveness of SE and NE in a power control problem to investigate the impact of the observation accuracy of the jammer, while \cite{cheng2022single} used  the hypergame framework to
discuss the robustness of SE strategies and NE strategies with misperception and deception.

\section{ Three-player	leader-follower security game}\label{s2}


We begin our study with a three-player  leader-follower security game, which refines a unified formulation from
several typical security games
 \cite{feng2015stealthy,wu2018secure,fang2018three}. 

%
%
%
Define the three-player security game by  $  \mathcal{G} = \{  \mathcal{X}\cup\mathcal{Y}\cup\mathcal{Z}, \Omega_{\mathcal{X}}\times\Omega_{\mathcal{Y}}\times\Omega_{\mathcal{Z}},  U_{\mathcal{X}}\cup U_{\mathcal{Y}}\cup U_{\mathcal{Z}} \}  $, {  where $ \mathcal{X} $, $ \mathcal{Y} $  and $ \mathcal{Z} $ are three players.}  Besides,
 $ \Omega_{\mathcal{X}}\subseteq \mathbb{R}  $, $ \Omega_{\mathcal{Y}}\subseteq \mathbb{R}  $, and $ \Omega_{\mathcal{Z}}\subseteq \mathbb{R}  $ are the strategy sets of players $ \mathcal{X} $,   $ \mathcal{Y} $, and $ \mathcal{Z} $,   respectively, 
  where  $  \Omega_{\mathcal{X}}=\{x|x_{\operatorname{min}}\leq x\leq x_{\operatorname{max}}\}  $, $  \Omega_{\mathcal{Y}}=\{y|y_{\operatorname{min}}\leq y\leq y_{\operatorname{max}}\}  $, and $  \Omega_{\mathcal{Z}}=\{z|z_{\operatorname{min}}\leq z\leq z_{\operatorname{max}}\}  $.  Moreover,  $U_{\mathcal{X}}: \Omega_{\mathcal{X}}\times \Omega_{\mathcal{Y}} \times \Omega_{\mathcal{Z}}\rightarrow \mathbb{R} $, $ U_{\mathcal{Y}}: \Omega_{\mathcal{X}}\times \Omega_{\mathcal{Y}} \times \Omega_{\mathcal{Z}}\rightarrow \mathbb{R} $, and $ U_{\mathcal{Z}}: \Omega_{\mathcal{X}}\times \Omega_{\mathcal{Y}} \times \Omega_{\mathcal{Z}}\rightarrow \mathbb{R} $ are the utility functions of players $ \mathcal{X} $,   $ \mathcal{Y} $, and $ \mathcal{Z} $,  respectively. Each player aims at maximizing its own
  utility.  Specifically, 
\vspace{-0.3cm}
\begin{subequations}\label{1a}
	\begin{align}
		& U_{\mathcal{X}}(x,y,z)=B(x)+f_{x}(y,z)x,\\\vspace{-0.3cm}
		& U_{\mathcal{Y}}(x,y,z)=f_{y1}(x,z)y+f_{y2}(x,z),\\\vspace{-0.3cm}
		& U_{\mathcal{Z}}(x,y,z)=f_z(x,y,z).
	\end{align}
\end{subequations}\vspace{-1.2cm}

	 One classical game model to reflect players’ strategic behaviors 
	is  the leader-follower  model 
	\cite{bacsar1998dynamic}, 
 and this hierarchical interplay reflected in the three-player  game  $ \mathcal{G} $ is a tri-level structure, 
 that is, $ \mathcal{X} $ at the top level, $ \mathcal{Y} $ at the middle level, and $ \mathcal{Z} $ at the  bottom  level.
{Players at high levels are called leaders, 
while players at low levels are called followers.}
The decision-making order  is as follows.
\begin{enumerate}[(1)]
	\item The top-level player $ \mathcal{X} $  first  determines  its strategy $ x $ to maximize its utility; 
	\item Observing $ x $, the
	middle-level player $ \mathcal{Y} $  then chooses its strategy $ y $ to  maximize its utility;
	\item Observing $ x $ and $ y $, the bottom-level player $ \mathcal{Z} $ finally adopts $ z $ to  maximize its utility.
\end{enumerate}	
Many security problems can be modeled by the  generalized  leader-follower game 	$ \mathcal{G} $. Here we introduce three practical  examples, which will be further investigated in Section \ref{s3}.

\textbf{Cooperative secure transmission}  
Consider a  secure transmission problem in a downlink  heterogeneous network \cite{wu2018secure,wu2016secure,bu2014interference}. 
The macro base station (MBS) within the network employs some small base stations (SBS) as jamming SBSs to jam the external eavesdropper for maximizing the secrecy rate, and each jamming SBS obtains the offloading service from the rest of the SBSs (called helping SBSs) in the cluster to satisfy the users. 
Set the number of jamming SBSs as one for simplification, as well as the helping SBSs.
 In  Fig. 1, the MBS is  player $ \mathcal{X} $,  the jamming SBS is  player  $ \mathcal{Y} $, and the helping SBS is  player $ \mathcal{Z} $.
\vspace{-0.5cm}
\begin{figure}[h]
	\centering	
	\includegraphics[scale=0.18]{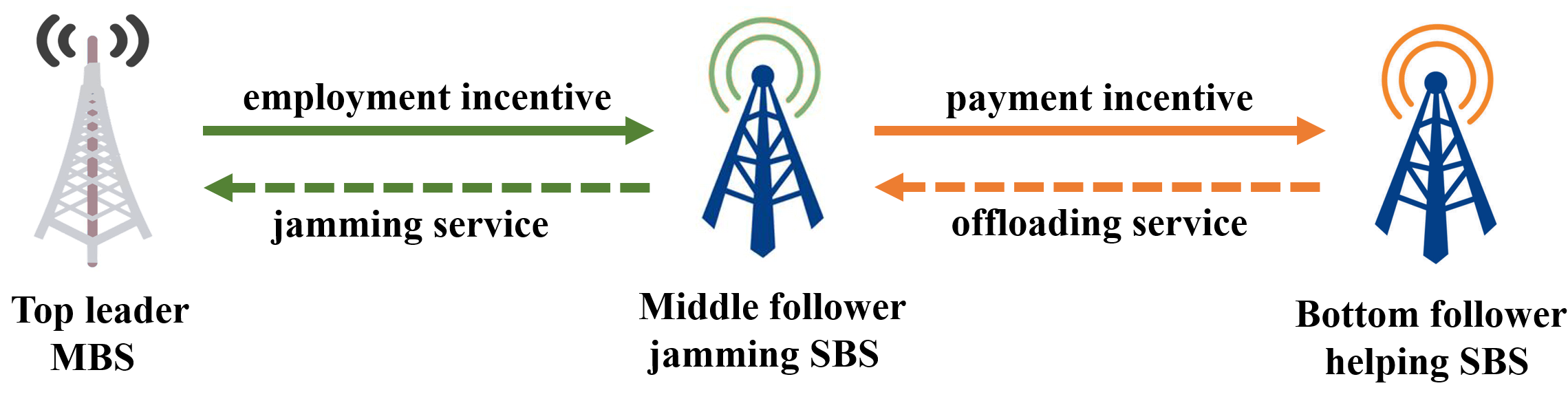}\vspace{-0.3cm}\\
	\caption{Tri-level 
		cooperative secure transmission problem. }
	\label{fig31}
	\vspace{-0.9cm}
\end{figure}

\textbf{
Adversarial cooperative communication} 
Consider  an adversarial cooperative communication system in a wireless network, in which the transmission from the source to the destination  is subject to an  eavesdropping attacks from an adversary \cite{fang2018three,tang2016combating,fang2017stackelberg}. 
To achieve cooperative communication and defend against eavesdropping attacks, the source  purchases the transmit powers from a selected relay, and this relay provides its relaying service for the source to obtain  benefits, while the  eavesdropper    broadcasts its jamming signal  to disrupt the transmission.  In  Fig. 2, the source, the relay, and the eavesdropper are  player $ \mathcal{X}  $,   $ \mathcal{Y} $, and $ \mathcal{Z} $, respectively.
\vspace{-0.5cm}
\begin{figure}[h]
	\centering	
	\includegraphics[scale=0.18]{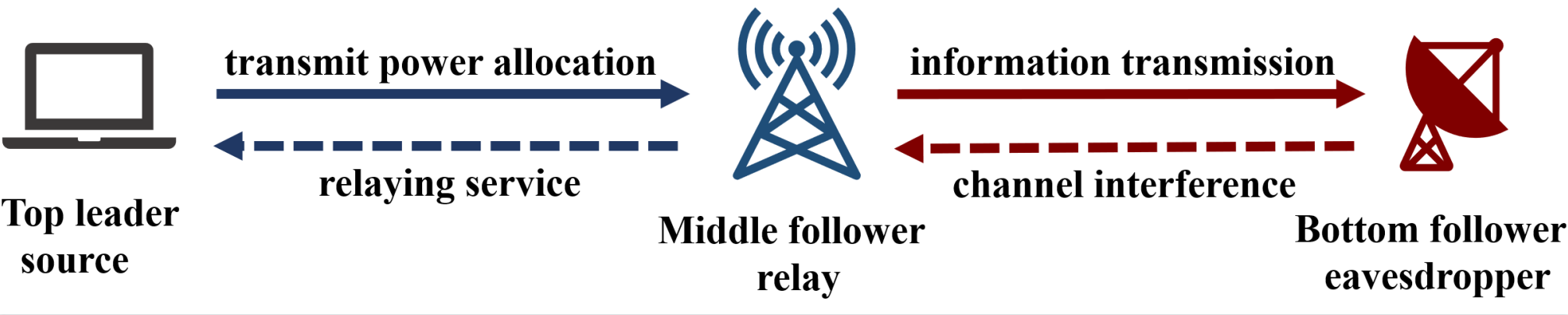}\vspace{-0.3cm}\\
	\caption{Tri-level defending against active eavesdropping attack problem. }
	\label{figw1}
	\vspace{-0.7cm}
\end{figure}

\textbf{Advanced persistent threat} 
 Consider an advanced persistent threat  with advanced attacks and insider threats \cite{feng2015stealthy,liu2021flipit}.
  The defender  and  the attacker take 	actions to gain control of the resource in the system, while an insider  with a privileged access to the  system  can monitor the defender's action and trade information to the attacker for its own profit. In  Fig. 3, the defender is  player $ \mathcal{X}  $,  the insider is  player $ \mathcal{Y} $, and the attacker  is  player  $ \mathcal{Z} $.
 \vspace{-0.5cm}
\begin{figure}[h]
	\centering	
	\includegraphics[scale=0.18]{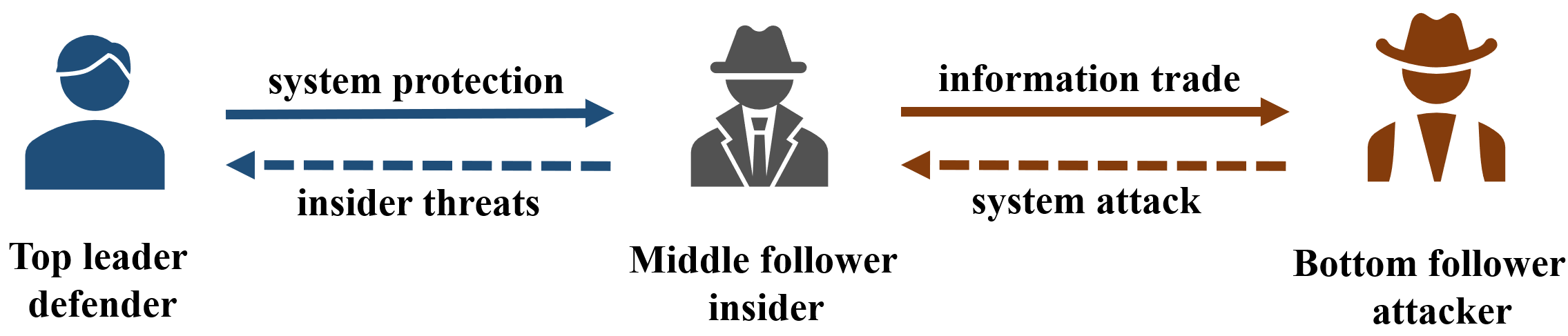}\vspace{-0.2cm}\\
	\caption{Tri-level APT problem. }
	\label{fig51}
	\vspace{-0.6cm}
\end{figure}

\begin{figure}[h]
 	\centering	
 	\includegraphics[scale=0.25]{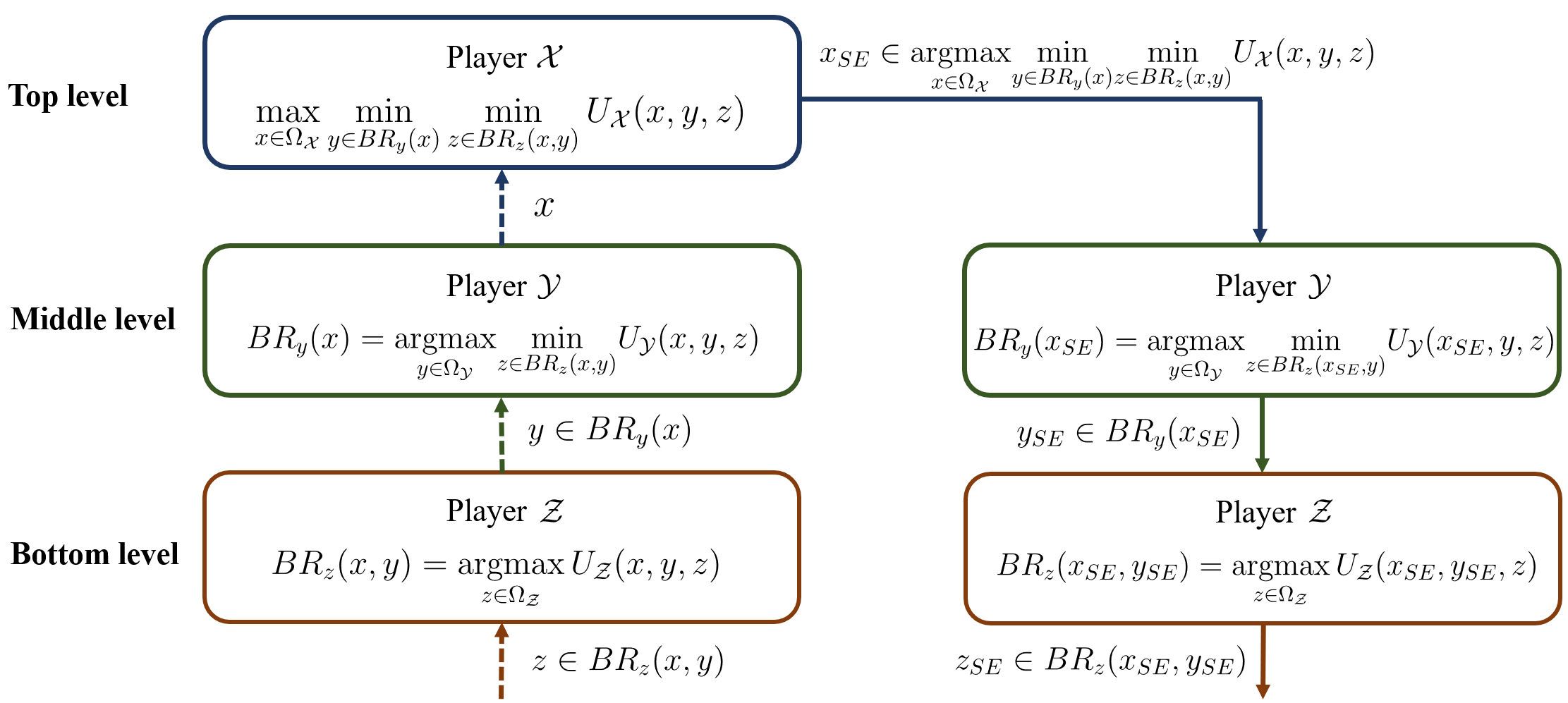}\vspace{-0.2cm}\\
 	\caption{The decision-making process of  three-player leader-follower security game $ \mathcal{G}$.}
 	\label{fig6}
 	\vspace{-0.6cm}
 \end{figure}

In the 
leader-follower scheme of $ \mathcal{G} $, 
	the low-level players adopt the   best response (BR) strategies based on
	the observed strategies of the high-level players, while
 the high-level players  compute their optimal strategies by considering low-level players.
Denote   $ \mathcal{Z} $'s best response to $ \mathcal{Y} $'s strategy $y$ and $ \mathcal{X} $'s strategy $x$ by
\vspace{-0.3cm}
$$
BR_{z}(x,y)= \{\omega\in \Omega_{\mathcal{Z}}: U_{\mathcal{Z}}(x,y,\omega )\geq U_{\mathcal{Z}}(x, y,z ), \forall z\in \Omega_{\mathcal{Z}} \}.  
\vspace{-0.5cm}
$$
Denote  $ \mathcal{Y} $'s best response to $ \mathcal{X} $'s strategy $x$ by
\vspace{-0.3cm}
$$
	BR_{y}(x)\!=\! \{\xi\in \Omega_{\mathcal{Y}}:\min_{z\in BR_{z}(x,\xi)}  U_{\mathcal{Y}}(x, \xi,z)\geq\min_{z\in BR_{z}(x,y )}  U_{\mathcal{Y}}(x, y,z), \forall y\in \Omega_{\mathcal{Y}}\}. 
	\vspace{-0.3cm}
$$
%
%
%
In this case, we introduce
the {Stackelberg equilibrium} (SE).
\begin{mydef}\label{g5}
For	the three-player leader-follower game $ \mathcal{G}$,  	
	a strategy profile $ (x_{SE}, y_{SE}, z_{SE}) $  is said to be an SE 
		 if 
		 \vspace{-0.3cm}
	$$
		\min_{y\in BR_{y}(x_{SE} )}\min_{ z\in BR_{z}( x_{SE},y)}
		 U_{\mathcal{X}}(x_{SE}, y,z) =\max_{x\in \Omega_{\mathcal{X}}} 
		 \min_{			y\in BR_{y}(x)} \min_{		z\in BR_{z}(x,y )} 
		U_{\mathcal{X}}(x, y,z),\vspace{-0.3cm}
	$$
	with $ y_{SE} \in BR_{y}(x_{SE} )$ and 
	   $ z_{SE} \in BR_{z}(x_{SE},y_{SE})$. 
\end{mydef}
Overall, the  conventional decision-making process of $ \!\mathcal{G}\!$ is given  as follows, as shown in
 Fig. $\!$\ref{fig6}.
 \begin{enumerate}[(1)]
	\item The  bottom-level player $ \mathcal{Z} $ solves  $	BR_{z}(x,y)=\underset{z\in\Omega_{\mathcal{Z}}}{\operatorname{argmax}} \, U_{\mathcal{Z}}(x, y,z) $ for any $x$ and $y$;
	\item The
	middle-level player $ \mathcal{Y} $ solves $  BR_{y}(x)= \underset{y\in\Omega_{\mathcal{Y}}}{\operatorname{argmax}} \underset{z\in BR_{z}(x,y)  }{\operatorname{min}} U_{\mathcal{Y}}(x, y,z) $ for any $x$;
	\item Then  $x_{SE}\!\in \!\underset{x\in  \Omega_{\mathcal{X}}}{\operatorname{argmax}}\!  \underset{y\in BR_{y}(x) }{\operatorname{min}} \underset{ z\in BR_{z}(x,y ) }{\operatorname{min}} 
\!	U_{\mathcal{X}}(x, y,z)$ is  an  SE strategy for  $ \mathcal{X} $;
	\item  The strategy $y_{SE}\in BR_{y}(x_{SE}) $ is an 
	SE strategy for $ \mathcal{Y} $;
	\item  The strategy $z_{SE}\in BR_{z}(x_{SE},y_{SE}) $ is an 
	SE strategy for $ \mathcal{Z} $;
	\item  The strategy profile $ (x_{SE}, y_{SE}, z_{SE}) $ constitutes  an SE of $ \mathcal{G}$.
\end{enumerate}	

 We give the following assumption for  game $  \mathcal{G} $.
%
\begin{assumption}
	$\ $
	\begin{enumerate}[(1)]
		\item $ B(x) \in \mathcal{C}^{1} $,  $ f_{z}(x,y,z) \in \mathcal{C}^{1} $   in $ z $, and $ f_{x}(y,z) \in \mathcal{C}^{1}$ in $ y $ and $ z $. For $ l=1,2, f_{yl}(x,y)   \in \mathcal{C}^{1} $ in $ x $ and $ y $. Moreover, $ BR_{y}(x) \in \mathcal{C}^{1} $ and $ BR_{z}(x,y)\in \mathcal{C}^{1} $ in $ x $ and $ y $.
		\item $ B(x) $ is concave in $ x $ and $ f_{z}(x,y,z)  $ is concave in $ z $.
	\end{enumerate}
\end{assumption}
	Assumption 1  guarantees the existence of SE \cite{bacsar1998dynamic,lucchetti1987existence}, which was also adopted in many practical
	security problems such as  secure transmission in the physical layer security \cite{wu2016secure},   
	IoT computational resource trading mechanism \cite{yang2020two}, APT defense problem \cite{feng2015stealthy},
	and cloud data computing issues \cite{liang2020investigating}.
	 The assumption about the continuous differentiability of $BR_{y}(x)$ and  $BR_{z}(x,y)$ 
	 guarantees that these best responses  are single-valued mappings rather than set-valued mappings	 
	 \cite{wu2016secure,feng2015stealthy,bacsar1998dynamic}, 
	implying that    $y_{SE}= BR_{y}(x_{SE}) $ for any given $x_{SE}$, and $z_{SE}= BR_{z}(x_{SE},y_{SE} )$ for   any given  $x_{SE}$ and  $y_{SE}$.
	 Moreover, Assumption 1 does not restrict the uniqueness of SE, 
	which is more general than those in some
	previous works \cite{wu2018secure,fang2018three,guruacharya2013hierarchical}. 
The following lemma,  whose proof is in Appendix A., verifies
the existence of an SE.
\begin{lemma}\label{l1}
	Under Assumption 1,  there exists an SE of $ \mathcal{G} $.
\end{lemma}

{ Although  the leader-follower scheme is indeed used in many  security scenarios, it may become invalid in practice. This is because the low-level player may lose the ability or interest to adopt the BR strategy and even ruin the leader-follower scheme, due to diverse factors such as the disturbance of the transmission
	environment  in cognitive
	radio network \cite{xiao2015anti},  the expensive
	surveillance cost of the defense strategy in the deployed infrastructure protection  \cite{an2013security},
	and the stealthy of the attack’s existence to avoid the fault detection \cite{xiao2018dynamic}. 
Hence, the high-level player may not
maintain
its dominant position, since its SE strategy is no
longer optimal  against the low-level player’s non-BR
strategy.
{
	In this view, 	the simultaneous-move game model may be an alternative option to reflect the practical situation,
	and the   best-known solution concept
 is the Nash equilibrium
 \cite{nash1951non}.
}}
\begin{mydef}
	For	the three-player leader-follower game $ \mathcal{G}$, 
	a strategy profile $ (x_{NE}, y_{NE}, z_{NE}) $ is said to be an NE 
	if 
	\vspace{-0.3cm}
	$$
	\begin{aligned}
		&x_{NE} \in \underset{x \in \Omega_{\mathcal{X}}}{\operatorname{argmax}}\; U_{\mathcal{X}}\left(x, y_{NE}, z_{NE}\right), \vspace{-0.3cm}\\
		&y_{NE} \in \underset{y\in \Omega_{\mathcal{Y}}}{\operatorname{argmax}}\; U_{\mathcal{Y}}\left(x_{NE}, y, z_{NE}\right), \vspace{-0.3cm}\\
			&z_{NE} \in \underset{z \in \Omega_{\mathcal{Z}}}{\operatorname{argmax}}\; U_{\mathcal{Z}}\left(x_{NE}, y_{NE}, z\right). \vspace{-0.3cm}
	\end{aligned}
$$
%
\end{mydef}
 {
 	It is acceptable for the high-level player
 	to accomplish such an NE when SE is not available, since no one can benefit from changing its
 	strategy unilaterally.} The following lemma  verifies
the existence of an NE,  whose proof is in Appendix A.
%
\begin{lemma}\label{l2}
	Under Assumption 1,	there exists an NE of game $ \mathcal{G} $. 
\end{lemma}
On this basis,  a high-level player has to decide which strategy should be adopted:
an SE under the leader-follower scheme or an NE under
the simultaneous-move scheme.
Clearly, players may  choose strategies with different  schemes, and  the derived conflict  may 
 bring a loss  for the high-level player's utility.
 Consider    two possible cases for an explanation. 
  One is that $\mathcal{X}$ adopts an SE strategy within the leader-follower scheme, while $\mathcal{Y}$ and $\mathcal{Z}$ act under the simultaneous-move scheme.
   In this way, the utility of $\mathcal{X}$  may be lower than  that when $\mathcal{X}$ acts under the simultaneous-move scheme, i.e.,
$U_{\mathcal{X}}(x_{SE}, y_{NE},z_{NE})\leq U_{\mathcal{X}}(x_{NE}, y_{NE},z_{NE}) 	$. 
   The other is that $\mathcal{X}$
   adopts an NE  strategy within the simultaneous-move scheme, while $\mathcal{Y}$ and $\mathcal{Z}$   act in the leader-follower scheme. 
This indicates that   $\mathcal{X}$'s utility may be lower than that when  $\mathcal{X}$ acts under the  leader-follower scheme, i.e., $ U_{\mathcal{X}}(x_{NE}, BR_{y}(x_{NE}),BR_{z}(x_{NE},BR_{y}(x_{NE})))\leq U_{\mathcal{X}}(x_{SE}, BR_{y}(x_{SE}),BR_{z}(x_{SE},BR_{y}(x_{SE})))$ $= U_{\mathcal{X}}(x_{SE}, y_{SE},z_{SE}) 	$. 

{However, it is worth mentioning that if an SE   is actually an  NE, 
then the high-level player will not meet these misgivings anymore. In such a case,
  the high-level player can be reassured to adopt an SE strategy since its utility is  the same as that of taking an NE strategy. 
Therefore, 
we expect to solve the following problem: 
\begin{itemize}
	\item 
	In what conditions,  SE coincide with  NE?
\end{itemize}
However, in many practical situations, SE and NE may not be  identical.
 If their difference is little,  the high-level player may still  adopt
an SE strategy.
Hence, we further ask the following question:
\begin{itemize}
	\item 
	If  the  coincidence condition cannot be guaranteed, 
	how to describe and measure the closeness between SE and NE?
\end{itemize}}

\section{SE coincident with NE}\label{s9}




In this section, we explore  the coincidence relationship between SE and NE in the three-player leader-follower security game $\mathcal{G}$.
Let  $ (x_{SE}, y_{SE}, z_{SE}) $ be an SE of $\mathcal{G}$.  It is clear that  $ (x_{SE}, y_{SE}, z_{SE}) $  can be equivalently described as $(x_{SE},BR_{y}(x_{SE}), BR_{z}(x_{SE},BR_{y}(x_{SE}) ))$. Obviously, player $ \mathcal{Z} $'s
SE strategy   
$ BR_{z}(x_{SE},BR_{y}$ $(x_{SE}) )=  BR_{z}(x_{SE},y_{SE})$ 
 becomes  an NE strategy
 when $ x_{SE}=x_{NE} $ and $y_{SE}=y_{NE}$. 
Hereupon, we  focus on the SE strategies for  $ \mathcal{X} $ and  $ \mathcal{Y} $ in the sequel.  

 In the leader-follower  decision-making process, by substituting $ z $ with $ BR_{z}(x,y) $,
 the composited utility function of  $ \mathcal{Y} $ is
$
\hat{U}_{\mathcal{Y}}(x,y)=f_{y1}(x,BR_{z}(x,y))y+f_{y2}(x,BR_{z}(x,y))
$.
The partial derivative of $ \hat{U}_{\mathcal{Y}} $ with regard to $ y $
is given as 
\vspace{-0.3cm}
$$
T_{1}(x,y)=\frac{\partial \hat{U}_{\mathcal{Y}}(x,y)}{\partial y}.
\vspace{-0.3cm}
$$
For any given $x=x_{SE}$, we have $T_{1}(x_{SE},y)$.
 Moreover, by  substituting  $ y $ with $ BR_{y}(x) $,
the composite utility function  of  $ \mathcal{X} $ is
$
	\hat{U}_{\mathcal{X}}(x)=B(x)+f_{x}(BR_{y}(x),BR_{z}(x,BR_{y}(x)))x.
$
Obviously, the  gradient of $ \hat{U}_{\mathcal{X}}$ is
\vspace{-0.3cm}
$$
T_{2}(x)=\frac{\partial \hat{U}_{\mathcal{X}}(x)}{\partial x}.
\vspace{-0.3cm}
$$
On the other hand, under the simultaneous-move scheme, $ \mathcal{Y} $ and  $ \mathcal{X} $ compute their optimal strategies based on the original  $ {U}_{\mathcal{Y}}(x,y,z) $ and $ {U}_{\mathcal{X}}(x,y,z) $, respectively.
 Take 
 \vspace{-0.3cm}
 $$T_{3}(x)=f_{y1}(x,BR_{z}(x,BR_{y}(x) )).
 \vspace{-0.3cm}
 $$ 
 The value of $T_{3}$ in $x_{SE}$ is equal to the partial derivative value of $ {U}_{\mathcal{Y}} $ with regard to $ y $  in   $ (x_{SE}, y_{SE}, z_{SE}) $. 
 Then let us take
 \vspace{-0.3cm}
 $$T_4(x)=  \frac{\partial B(x)}{\partial x}+f_{x}(y_{SE},z_{SE}),
 \vspace{-0.3cm}
 $$ 
which can be regarded as the partial derivative value of $ {U}_{\mathcal{X}} $ in $ x $ for given $y=y_{SE}$, $z=z_{SE}$. 

Denote $\delta(\cdot)$ as the neighbourhood of one point, $\delta_{-}\!(\cdot)$ as the left neighbourhood and  $\delta_{+}\!(\cdot)$ as the right neighbourhood. 
The following assumption is about the local  monotonicity of  utility functions, which is more relaxed than the global monotonicity and strict monotonicity \cite{zhang2021bayesian,marchesi2018leadership}.
\begin{assumption}
	For $  y \in \Omega_{\mathcal{Y}}$, there exist $ \delta_{-}(y)$ and $ \delta_{+}(y) $ such that $\hat{U}_{\mathcal{Y}} $ is monotone in $ \delta_{-}(y)\cap \Omega_{\mathcal{Y}} $ and $ \delta_{+}(y)\cap \Omega_{\mathcal{Y}} $. For $ x \in \Omega_{\mathcal{X}}$, there exist $ \delta_{-}(x)$ and $\delta_{+}(x)$ such that $\hat{U}_{\mathcal{X}} $ is monotone in $ \delta_{-}(x)\cap \Omega_{\mathcal{X}} $ and $ \delta_{+}(x)\cap \Omega_{\mathcal{X}} $.
\end{assumption}
%
 
{In the following, 
we  provide a necessary and sufficient condition for the case  that each SE is an NE in the three-player leader-follower security game $\mathcal{G}$,  whose proof is in Appendix B.}



\begin{mythm}\label{t1}
	Under Assumptions 1 and 2, any  SE is an NE  if and only if there exists $ \delta(x_{SE})$ and $ \delta(y_{SE}) $ such that 
	
	\begin{enumerate}[(i)]
		\item $T_{1}(x_{SE},y)\cdot T_{3}(x_{SE}) \geq0 $ for $  y \in \delta(y_{SE})\cap \Omega_{\mathcal{Y}}$;
		\item $ T_{4}(x_{SE})=0 $ or $ T_{2}(x)\cdot T_{4}(x) >0 $ for $  x \in \delta(x_{SE})\cap \operatorname{rint}(\Omega_{\mathcal{X}}) $.
	\end{enumerate}	
		
%
	
	%
\end{mythm}

%



Theorem \ref{t1} provides the coincidence condition to connect SE and NE. 
If the condition is satisfied, then the high-level players can get rid of the strategy selection dilemma, as they can safely adopt SE strategies. 
From the sufficiency perspective,   the condition develops an approach to seek an SE  that is exactly an NE. The approach covers all possible cases in which each player's SE strategy may be  the boundary point or interior point of its strategy set, so that we can  directly confirm whether the set of SE  is a subset of NE.
On the other hand, from the necessity perspective,   the condition  provides a criterion  to verify whether an obtained SE is an NE.
  The computation is not complicated because  the partial derivatives therein may usually be zero \cite{fang2018three,wang2015pricing},
  and  are merely related to the local information of strategy sets. 

In fact, this equilibrium coincidence analysis  is important and can be employed  in many practical security scenarios.
    	 In  adversarial cooperative communication  issues \cite{fang2018three,tang2016combating}, the coincidence condition   becomes an inequality merely depending on
    	 the  strategy of the source  $\mathcal{X}$ and  the parameters of different channel gains. In APT problems with insider threats \cite{feng2015stealthy,liu2021flipit}, the condition is embodied as inequalities related to the defense and attack cost parameters. 
    	 Readers can see Section \ref{s3} for more details.

Moreover, in the case when 
 the SE is a unique solution,
  we have the following  result,
 whose proof is in Appendix  C.
\begin{corollary}\label{c1}
	Under Assumptions 1 and 2 and provided that the SE is unique, the   SE is  an NE  if and only if there exist $ \delta(x_{SE})$ and $ \delta(y_{SE}) $ such that  	
		\begin{enumerate}[(i)]
		\item  $T_{1}(x_{SE},y)\cdot T_{3}(x_{SE}) \geq0 $ for $  y \in \delta(y_{SE})\cap \Omega_{\mathcal{Y}}$;
		\item $ T_{2}(x)\cdot T_{4}(x) \geq 0 $ for $  x \in \delta(x_{SE})\cap \Omega_{\mathcal{X}} $.  
	\end{enumerate}	
%
\end{corollary}

{Moreover, 
when not all SE  are NE, we turn our attention to whether there exists an SE that is an NE, and  provide a  condition for the case that at least one SE is an NE in the following result,
whose proof is shown in Appendix D.}

\begin{mythm}\label{c2}
Under Assumptions 1 and 2, at least one  SE is an NE  if and only if there exists $  \delta(y_{SE}) $ and $  \delta(x_{SE}) $ such that  
\begin{enumerate}[(i)]
	\item	  $T_{1}(x_{SE},y)\cdot T_{3}(x_{SE}) \geq0 $ for $  y \in \delta(y_{SE})\cap \Omega_{\mathcal{Y}}$;	
	\item $  T_{4}(x)\cdot \operatorname{sgn}(x-x_{SE}) \leq 0 $ for $  x \in \delta(x_{SE})\cap \Omega_{\mathcal{X}}$.
\end{enumerate}
\end{mythm}
{Theorem \ref{c2} shows a necessary and sufficient condition   for the existence of an SE that is an NE. Different from the discussion of the entire  SE set in Theorem \ref{t1}, the  analysis in Theorem \ref{c2} focuses on the specific  SE. In this way, the high-level players can employ this condition to exactly find out  a satisfactory equilibrium   and adopt the
corresponding SE strategy.}
%
\section{SE close to NE}\label{s8}


In reality, the  coincidence between SE and NE may not always happen. {
Therefore, we expect to find a way to  measure the difference
	between SE and NE
	 so as to help high-level players make a reasonable decision.

{
Here, we  employ  the Hausdorff metric to describe the closeness of SE and NE. }
Define the Hausdorff metric of two sets $ A,B\subseteq  \mathbb{R}^{n} $ by 
\vspace{-0.3cm}
$$ 
H(A,B) = \max\{\sup\limits_{a\in A}\operatorname{dist}(a,B),\sup\limits_{b\in B}\operatorname{dist}(b,A) \}.
\vspace{-0.3cm}
$$
%
 Denote $\Xi_{SE}$ as the set of  SE strategy profile $(x_{SE},y_{SE},z_{SE})$, and $\Xi_{NE}$ as the set of  NE strategy profile $(x_{NE},y_{NE},z_{NE})$. 
For any SE strategy profile $p^*\in \Xi_{SE} $ with  $p^*\triangleq(x_{SE},y_{SE},z_{SE})$, take the operator $T_2(\cdot)$ on the element $x_{SE}$ from $p^*$ and denote $\Pi_{x_{SE}}$ as the image set of $T_2(x_{SE})$.
  Also,  take the operator $T_1(\cdot)$ on the pair $(x_{SE}, y_{SE})$ from $p^*$
 and denote $\Pi_{y_{SE}}$ as the image
set of $T_1(x_{SE},y_{SE})$.
Similarly,  for any NE strategy profile $q^*\in \Xi_{NE} $ with  $q^*\triangleq(x_{NE},y_{NE},z_{NE})$,  take the operator $T_2(\cdot)$ on the element $x_{NE}$ from $q^*$ and denote $\Pi_{x_{NE}}$ as the image set of $T_2(x_{NE})$.
 Also,  take the operation $T_1(\cdot)$ on the pair $(x_{SE}, y_{NE})$, where $x_{SE}$ is chosen from any given $p^{*}$ and $y_{NE}$ is choosen from any given $q^{*}$.
 Denote $\Pi_{y_{SE}}$ as the image
 set of $T_1(x_{SE},y_{NE})$.

Then the closeness of  SE  and  NE  is estimated in the following result, whose proof is in Appendix E.

\begin{mythm}\label{c7}
	Under Assumption 1, if there exist  constants $\kappa_1>0$, and $\kappa_2>0$ such that $\hat{U}_{\mathcal{Y}}$ is $\kappa_1$-strongly concave in $y$ and $\hat{U}_{\mathcal{X}}$ is $\kappa_2$-strongly concave in $x$, then
	with  	
 $\max\{ H(\Pi_{x_{SE}},\Pi_{x_{NE}}),H(\Pi_{y_{SE}},$ $\Pi_{y_{NE}}) \}<\eta$, we have $H(\Xi_{SE},\Xi_{NE})<\frac{(1+l)(\kappa_1+\kappa_2)}{\kappa_1\kappa_2}\eta $. 
\end{mythm}

Theorem \ref{c7} provides the closeness of SE and NE by giving an upper bound of the distance between their corresponding sets.
In addition to the  Lipschitz constant $l$, the strong concavity constants $\kappa_1$ and $\kappa_2$,
 the  upper bound of the Hausdorff metric $H\!(\Xi_{SE},\Xi_{NE})$ is mainly affected by  the maximal value between $H\!(\Pi_{x_{SE}},\Pi_{x_{NE}})$ and $H\!(\Pi_{y_{SE}},\Pi_{y_{NE}})$.  Regarding this maximal value
 as a
 perturbation,  it is  clear that a lower  perturbation yields a lower bound.
If  the bound  is low enough, then  SE can be regarded as close to NE.
This indicates that high-level players  can still be reassured to adopt the SE strategy, as the brought deviations in their utilities are
 tolerable.
 
 {Additionally, when  both the SE and the NE are  unique solutions in some security issues, we can obtain the upper bound of the distance between these two equilibrium points in  the following result, whose
 proof can be easily modified from Theorem \ref{c7}.}
 \begin{corollary}\label{co1}
 	Under Assumption 1 with that both the SE and the NE are unique, if there exist  constants $\kappa_1>0$, and $\kappa_2>0$ such that $\hat{U}_{\mathcal{Y}}$ is $\kappa_1$-strongly concave in $y$ and $\hat{U}_{\mathcal{X}}$ is $\kappa_2$-strongly concave in $x$, then
 	with  	
 	$\|T_1(x_{SE},y_{SE})-T_1(x_{SE},y_{NE}) \|<\eta_{1}$ and $\|T_2(x_{SE})-T_2(x_{NE}) \|<\eta_{2}$, we have $\|p^*-q^*\|<(1+l)(\frac{\eta_1}{\kappa_1}+\frac{\eta_2}{\kappa_2})$. 
 \end{corollary}

\section{Applications }\label{s3}


In this section,  we demonstrate  our theoretical results in several important  security games (introduced in Section \ref{s2}), and  further illustrate the equilibria relationship for different scenarios.  


%
%

	\subsection{{Adversarial cooperative communication with eavesdropping attack}}

Consider a  security issue on  defending against eavesdropping attacks in the cooperative
communication system, consisting of a   primary source (player $\mathcal{X}$), a relay (player $\mathcal{Y}$), and an eavesdropper  (player $\mathcal{Z}$)
\cite{fang2018three,tang2016combating,fang2017coordinated}.  {
	 The source  first decides the  transmit  power purchased from the selected relay  to defend against the eavesdropping attacks, and then  the relay decides the price of the unit power,
while the eavesdropper  finally decides its jamming power to disrupt the legitimate transmission based on the channel information and behavioral information of the relay and the source.}
Denote $ x $ as the  amount of the purchased transmit
power, $ y $  as the   price set by the relay, and $ z $ as the  amount of the jamming power.  
{Referring to \cite{fang2018three,tang2016combating,fang2017coordinated}, the three-player game 	 is modeled as }
\vspace{-0.1cm}
\begin{equation*}\label{f4}
	\begin{aligned}
		\max_{x\in [x_{\operatorname{min}},x_{\operatorname{max}}]} &U_{\mathcal{X}}(x,y,z)= \frac{d_{1}|h_{rd}|^2 x}{\eta+|h_{ed}|^2z} -d_{4}xy ,\vspace{-0.3cm}\\		
		\max_{y\in [y_{\operatorname{min}},y_{\operatorname{max}}]} &U_{\mathcal{Y}}(x,y,z)=xy-d_{3}x,\vspace{-0.3cm}	\\
		\max_{z\in [z_{\operatorname{min}},z_{\operatorname{max}}]} &U_{\mathcal{Z}}(x,y,z)= -\log_{2}(\frac{\left|h_{r d}\right|^{2} x+\eta+\left|h_{e d}\right|^{2} z}{\eta+\left|h_{e d}\right|^{2} z}) -d_{2}z, 
	\end{aligned}
\end{equation*}
where $h_{rd}$ and $h_{ed}$  are the respective channel gains of the  relay-destination link and eavesdropper-destination link with $h_{rd}, h_{ed}\in[0,1]$, 
$\eta$ indicates the background noise on the channel, 
$d_{1}$ is the gain coefficient, and $d_{i}$ are the cost coefficients for $i=\{2,3,4\}$. { 
In this model, we denote
} 
 $U_{\mathcal{X}}$ as the benefits of the source  from the secure cooperative transmission with $f_{x}(y,z)x=\frac{d_{1}|h_{rd}|^2 x}{\eta+|h_{ed}|^2z} -d_{4}xy$,  
$U_{\mathcal{Y}}$ as the combination of the  relaying payment given by the source and the  relay transmission cost with
 $f_{y1}(x,z)y=xy$ and $f_{y2}(x,z)=-d_{3}x$, and
$U_{\mathcal{Z}}$  as the benefit of the eavesdropper from reducing secrecy capacity.


 {Due to the expensive  cost of eavesdropping or  selfish concerns for own benefits \cite{fang2018three,tang2016combating}, the  eavesdropper or the relay may lose  interest to obtain the whole transmission information  and break down the leader-follower scheme. Thus, the cooperative communication may not be guaranteed and the source's utility may suffer a loss.}
To reassure the source, 
 we  investigate the coincidence  between SE and NE in this three-player game.
It can be derived that
\vspace{-0.3cm}
$$ 	
\begin{aligned}
	&T_{1}(x,y)=x,\quad\quad T_{2}(x)=  \frac{d_{1}|h_{rd}|^{4}|h_{ed}|^2}{\phi_{x}(\ln2d_{2}|h_{rd}|^{4}x+2|h_{rd}|^2|h_{ed}|^2-2\ln2d_{2}|h_{rd}|^2\phi_{x})}-d_{4}y_{SE},\\
		&T_{3}(x)= x,\quad\quad\quad 
		T_{4}(x)= \frac{d_{1}|h_{rd}|^2}{\eta+|h_{ed}|^2z_{SE}}-d_{4}y_{SE}\vspace{-0.3cm} 
\end{aligned}
$$
where 	 $\phi_{x}=\sqrt{\frac{|h_{rd}|^{4}x^{2}}{4}+\frac{|h_{rd}|^2|h_{ed}|^2x}{\ln2 \; d_{2}}}
$.
Obviously,  under Assumptions 1 and 2,  there exists $  \delta(y_{SE}) $, $  \delta(x_{SE}) $ such that  $T_{1}(x_{SE},y)\cdot T_{3}(x_{SE}) \geq0, $ for $  y \!\in\! \delta(y_{SE})\!\cap\! \Omega_{\mathcal{Y}}$ and $T_{4}(x)>0$ for $  x \in \delta(x_{SE})\cap \operatorname{rint} (\Omega_{\mathcal{X}}) $. Thus, the  coincidence condition in Theorem \ref{t1} is simplified to analyze $T_{2}(x)$. Due to $y_{SE}\!=\!y_{\operatorname{max}}$ \cite{fang2018three}, any SE is an NE if and only if there exists  $  \delta(x_{SE}) $ such that
\vspace{-0.3cm}
\begin{equation}\label{f44}
\phi_{x}(|h_{rd}|^{4}d_{2}x+2|h_{rd}|^2|h_{ed}|^2-2|h_{rd}|^2d_{2}\phi_{x})<\frac{d_{1}|h_{rd}|^{4}|h_{ed}|^2}{2d_{2}d_{4}y_{\operatorname{max}}}, \quad   \forall x \in \delta(x_{SE})\cap \operatorname{rint}(\Omega_{\mathcal{X}}).\vspace{-0.3cm}
\end{equation}
It follows from \eqref{f44} that the coincidence condition in this problem is transformed into an inequality merely depending on the source's strategy and the parameters of different channel gains. 
Moreover,	 the channel gain of relay-destination link $|h_{rd}|^{2}$ in \eqref{f44}
has a large impact on the players' strategies and their utilities and		 may vary significantly due to the change of wireless networks \cite{fang2017coordinated}.
} Thus, we
set  $|h_{rd}|^2=0.2$
and    $|h_{rd}|^2=0.7$ 
as two  environment settings herein. Then we consider   three  strategy profiles: each player  chooses the SE strategy $(x_{SE}, y_{SE},z_{SE})$; the  source takes the SE strategy while the relay and the eavesdropper  adopt NE strategies  $(x_{SE}, y_{NE},z_{NE})$; each player  chooses the NE {strategy} $(x_{NE}, y_{NE},z_{NE})$. 
 With these strategy profiles, Fig. \ref{fig17}  shows the utilities of the source  in different settings. { 
 	In  Fig. \ref{fig17}(a),
 	 the SE does not coincide with the NE. 
If the source insists on the SE strategy, its utility may decrease from ideal case 1 to case 2 since the relay may not forward packets and the eavesdropper may become passive, which makes the leader-follower scheme invalid.  Adopting the NE strategy is an acceptable choice for the source, as all players can still reach the equilibrium even when  the cooperative communication may not be guaranteed, and the source's utility in case 2 is higher than  that in  case 3.
 Thus, the source needs to make a trade-off between the SE and NE strategies.
However, in Fig. \ref{fig17}(b), 
the SE is indeed the NE.
  It reflects that the source can be reassured to adopt the SE strategy and  its utility does not change in each case.
{Thus,  once the coincidence condition \eqref{f44} is satisfied,  
	there is no  strategy selection dilemma for the source. Regardless of whether the relay or the  eavesdropper can obtain the	whole transmission information,
	  the  system security    can be guaranteed and
the secure transmission performance can be	improved  \cite{fang2018three,tang2016combating,fang2017coordinated}. Moreover, we establish the coincidence relationship of SE and NE by analyzing the complicated interplay among multiple hierarchies in a three-player problem,  which is beyond the consideration of models merely involving two players \cite{xiao2015anti,tang2016combating}.}
\vspace{-0.5cm}
\begin{figure}[h]
	\hspace{-1.5cm}
	\centering	
	\begin{subfigure}[t]{0.55\linewidth}
		\centering
		\includegraphics[width=6cm]{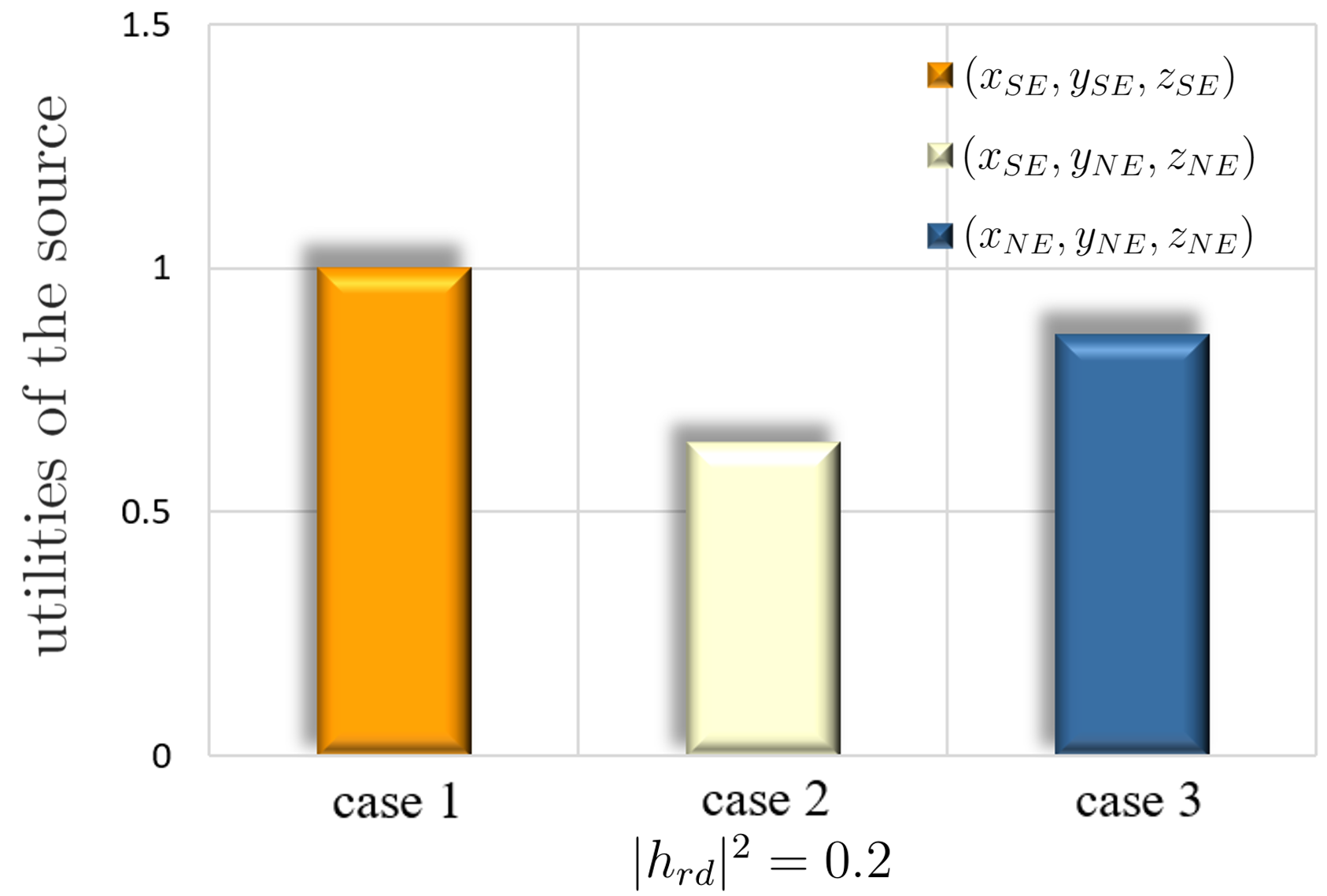}
	\end{subfigure}%
	\hspace{-1.2cm}
	\begin{subfigure}[t]{0.55\linewidth}
		\centering
		\includegraphics[width=6cm]{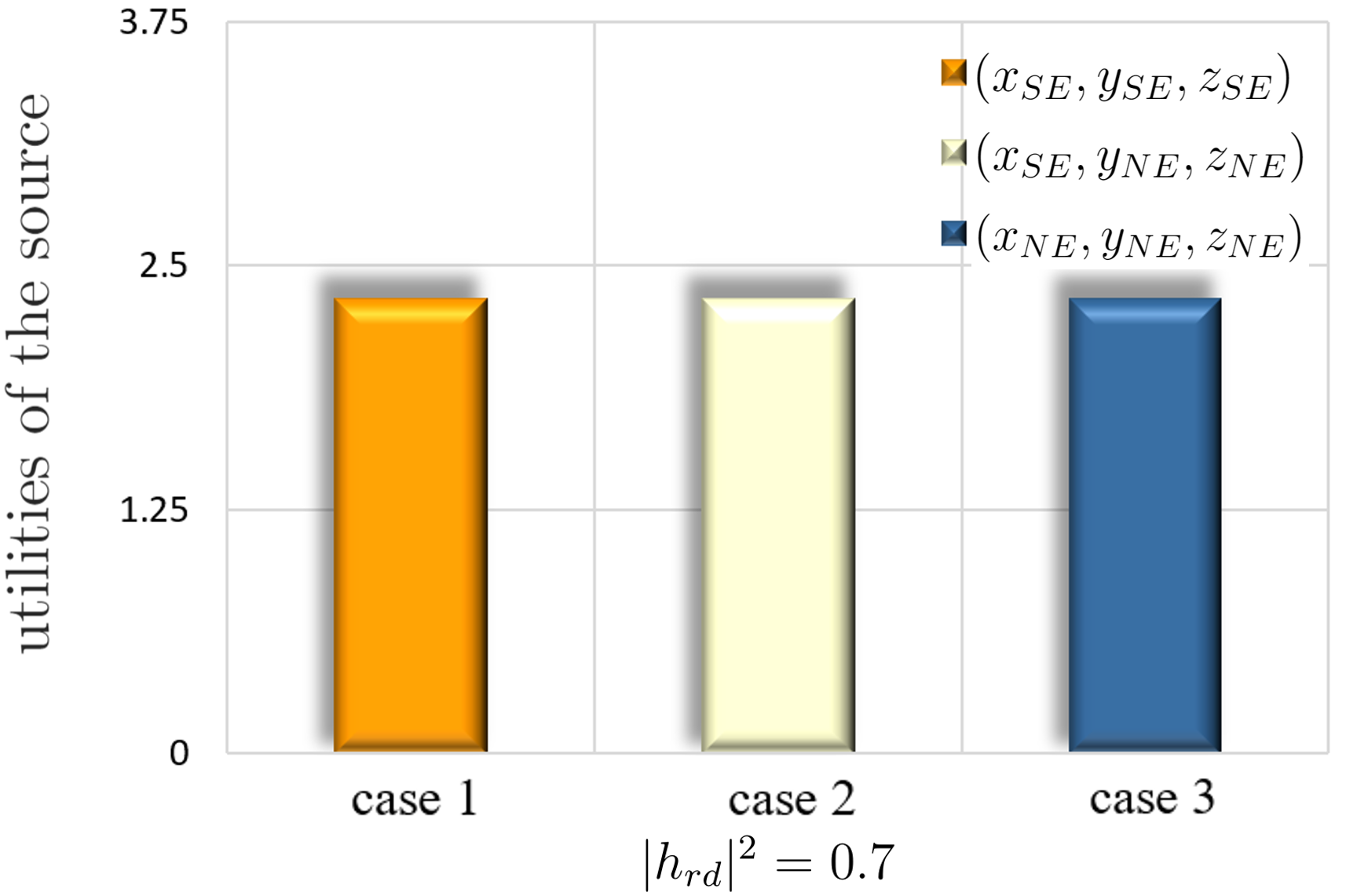}
	\end{subfigure}%
	\centering
	\vspace{-0.1cm}
	\caption{Utilities of the source under different strategy profiles.} 
	\label{fig17}
	\vspace{-0.65cm}
\end{figure}

\subsection{Advanced persistent threats (APT) with  insider threats }


Consider a three-player APT game with advanced attacks and insider threats
in cyber security, consisting of a   defender (player $\mathcal{X}$), an insider (player $\mathcal{Y}$), and an attacker  (player $\mathcal{Z}$) \cite{feng2015stealthy,feng2016stealthy,liu2021flipit}.   {After the defender first  determines its defense rate, the insider  determines the amount of the traded inside information to the attacker, and finally, the attacker  chooses its attack rate. }
%
%
%
Denote $ x $ as the  defense rate of the defender, $ y $ as the amount of the traded  information of the insider, and $ z $ as the  attack rate of the attacker. 
{Referring to \cite{feng2015stealthy},}  this game  is designed as 
%
\vspace{-0.35cm}
		\begin{align*}
			\max_{x\in [x_{\operatorname{min}}, x_{\operatorname{max}}]}\; &U_{\mathcal{X}}(x,y,z)=\frac{x}{2z} -C_{\mathcal{D} }x,\\\vspace{-0.35cm}
		\max_{y\in [y_{\operatorname{min}}, y_{\operatorname{max}}]}\;& U_{\mathcal{Y}}(x,y,z)=\rho\frac{x}{2z} + y,\\\vspace{-0.35cm}
		\max_{z\in [z_{\operatorname{min}}, z_{\operatorname{max}}]}\; &U_{\mathcal{Z}}(x,y,z)=1-\frac{x}{2z}-C_{\mathcal{A} }(1-y)^{2}z -y,\vspace{-0.35cm}
	\end{align*}
where $ C_{\mathcal{D} } $ is the cost for each defense action, $ \rho<1 $ is the constant denoting the insider's proportion in
the system with the upper bound $  y_{\operatorname{max}}\leq \rho $  to restrict the capability of the insider, and $ C_{\mathcal{A} } $ is the cost for each attack action.
 The first term in $U_{\mathcal{X}}$ is  the gain from the protected system while the second term is the cost of recapturing the compromised
 resources, where  $B(x)=-C_{\mathcal{D} }x$ and  $f_{x}(y,z)x=\frac{x}{2z}$. 
   The first term in  $U_{\mathcal{Y}}$ represents the profit of selling inside
  information, while the second term is the profit from the protected system, where $f_{y1}(x,z)y=y$ and $f_{y2}(x,z)=\rho\frac{x}{2z}$.  
 Moreover,  the first two terms in $U_{\mathcal{Z}}$ present
 the benefit from the compromised system resource, and the third term denotes  the cost of launching attacks, while the last term means the cost of  purchasing information from the insider.

 Accordingly, 
 \vspace{-0.2cm}
$$ 	
\begin{aligned}
	&T_{1}(x,y)=1-\sqrt{\frac{\rho^{2} C_{A}x}{2}},\quad T_{2}(x)= (1-\rho)\sqrt{\frac{C_{A}}{8x}}-C_{D},\quad T_{3}(x)=  1, \quad T_{4}(x)=\frac{1}{2z_{SE}}-C_{D}.\vspace{-0.2cm}
\end{aligned}
$$
In this way, 
we obtain that under Assumptions 1 and 2,  any SE is an NE  if and only if 
\vspace{-0.3cm} 
\begin{equation}\label{er3}
	\frac{C_{A}}{C_{D}^{2}}\geq  \frac{8x_{\operatorname{max}}}{(1-y_{\operatorname{max}})^{2}}\quad   \text{or} \quad \frac{C_{A}}{C_{D}^{2}}\leq  \frac{2x_{\operatorname{min}}}{(1-y_{\operatorname{min}})^{2}}.\vspace{-0.3cm}    
\end{equation}
From    (\ref{er3}),  the coincidence between SE and NE is mainly affected by the attack cost parameter $\!C_{A}$ and the defense cost parameter $C_{D}$. {The configuration of these two parameters plays an important role in APT issues, and affects the utility of players \cite{feng2015stealthy,liu2021flipit}.}
Set $C_{A}\!\!\in\![0.44,1.25]$ and $C_{D}\!\in\![0.15,0.55]$.
Fig. \ref{fig15}(a) first provides the coincidence ratios between SE and NE under different settings of $C_{A}$ and $C_{D}$.
Clearly,  the ratio  varies in different ranges with the  changes of $C_{A}$ and $C_{D}$,  and it increases  when  $\!{C_{A}}/{C_{D}}\!$  becomes large. Moreover,
if $C_{A}$ and $C_{D}$ correspond to the dark areas, then SE coincide with NE, and the high-level players can safely
take  SE strategies.     
On the other hand, 
 Fig. \ref{fig15}(b)  shows the defender's utilities according to different parameter values in Fig. \ref{fig15}(a). The blue line describes  the defender's utility  with SE  strategies $\!(x_{SE}, y_{SE},z_{SE})\!$, while the red line describes the defender's utility with NE  strategies $\!(x_{NE}, y_{NE},z_{NE})\!$.
   As can be seen from each subfigure of Fig. \ref{fig15}(b), a smaller $C_{D}$ means that the defender can protect the system with less cost, which corresponds to the higher utility; a
 	larger $C_{A}$ means that attackers need to take more cost to compromise the resource system, which also yields the defender's higher utility \cite{feng2015stealthy}. More importantly, 	when  $C_{A}$ and $C_{D}$ satisfy  condition  (\ref{er3}), the defender's utility in the SE strategy is the same as that in the NE strategy.  {  This indicates that the defender  can achieve efficient defense when  $C_{A}$ and $C_{D}$ are maintained in an acceptable range,  even facing some misgivings brought by stealthy  attacks or  unknown insider trading  in some APT issues \cite{feng2015stealthy,feng2016stealthy,chen2022defense}, including the three-player problem that only discusses NE \cite{chen2022defense}.
 	} 		
 		\vspace{-0.5cm}
 	\begin{figure}[h]
 	\centering	
 	\begin{subfigure}[c]{0.5\linewidth}
 		\centering
 		\includegraphics[width=6.6cm]{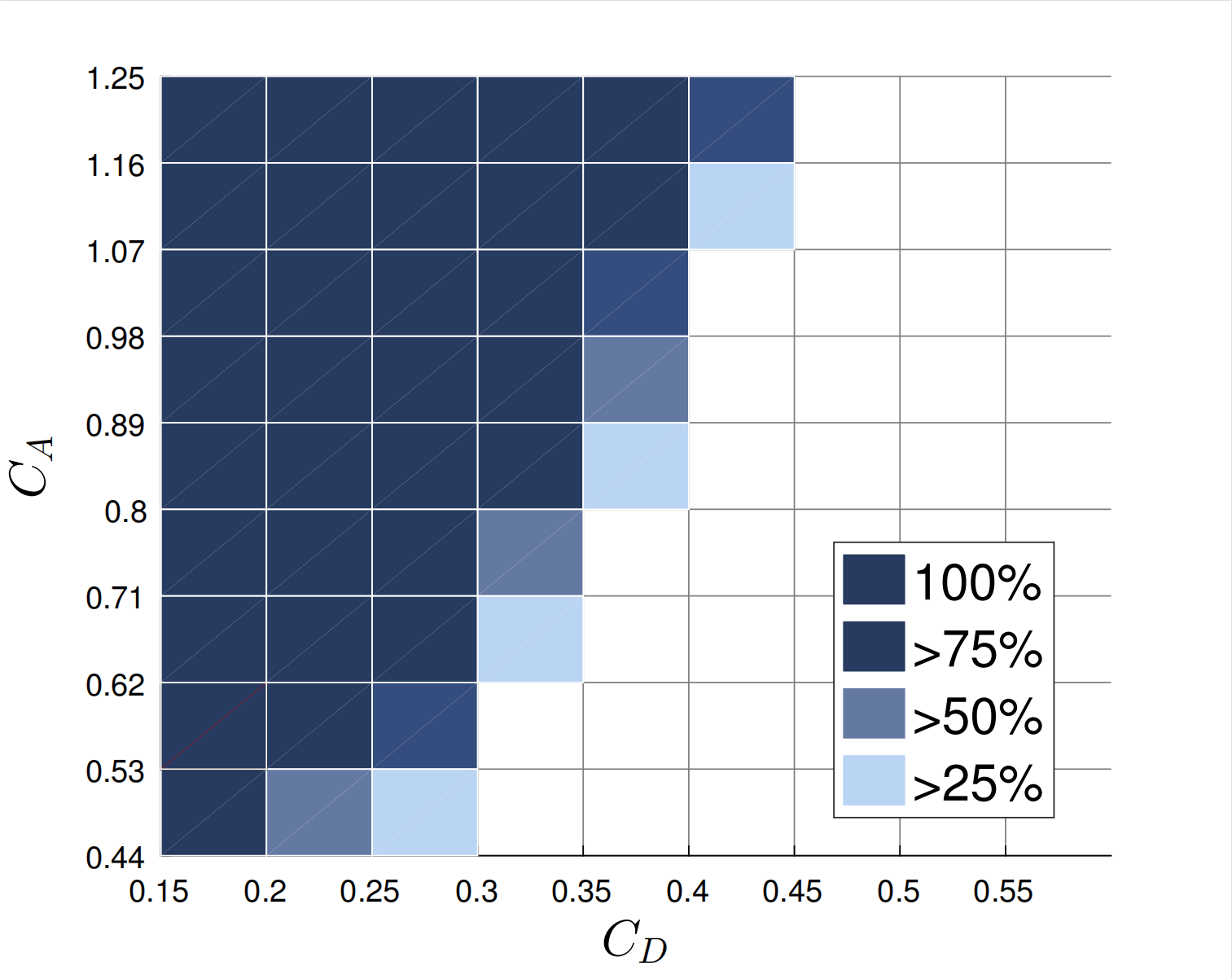}
 		\vspace{-0.08cm}
 		\caption{Coincidence ratios  between SE and NE}
 	\end{subfigure}%
 	\begin{subfigure}[c]{0.5\linewidth}
 		\begin{subfigure}[c]{0.45\linewidth}
 			\centering
 			\includegraphics[width=3.1cm]{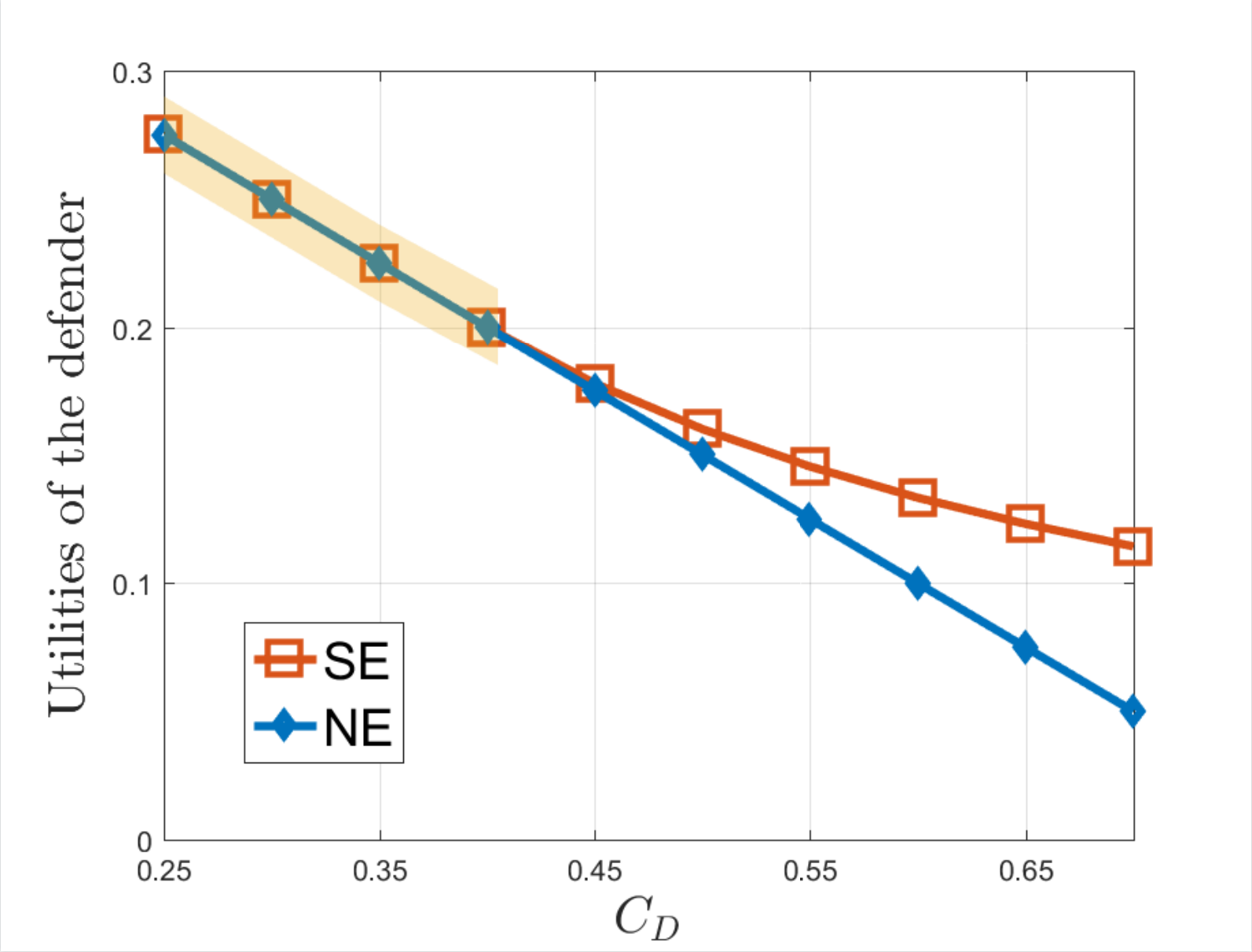}\\
 			\includegraphics[width=3.1cm]{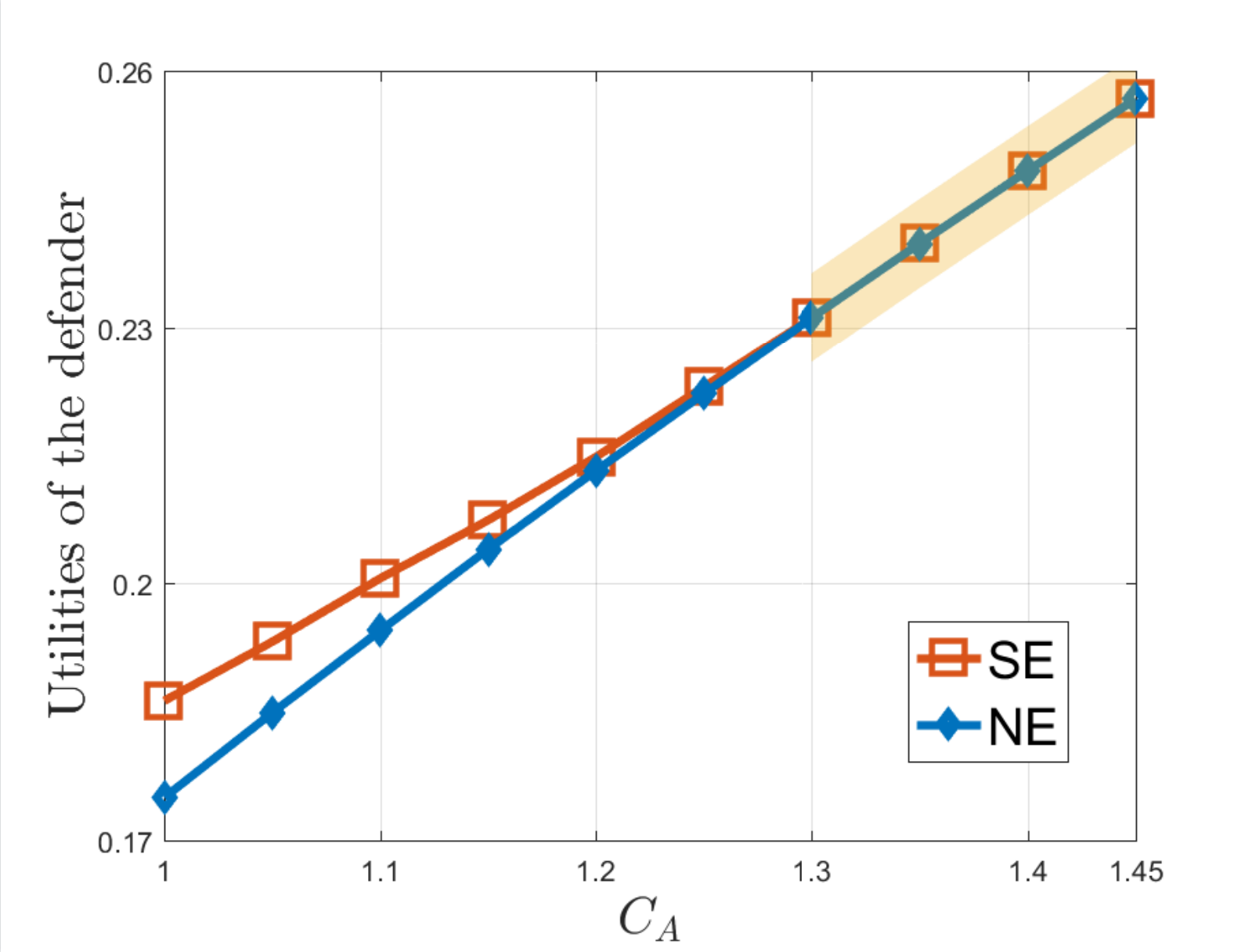}
 		\end{subfigure}%
 		\begin{subfigure}[c]{0.45\linewidth}
 			\centering
 			\includegraphics[width=3.1cm]{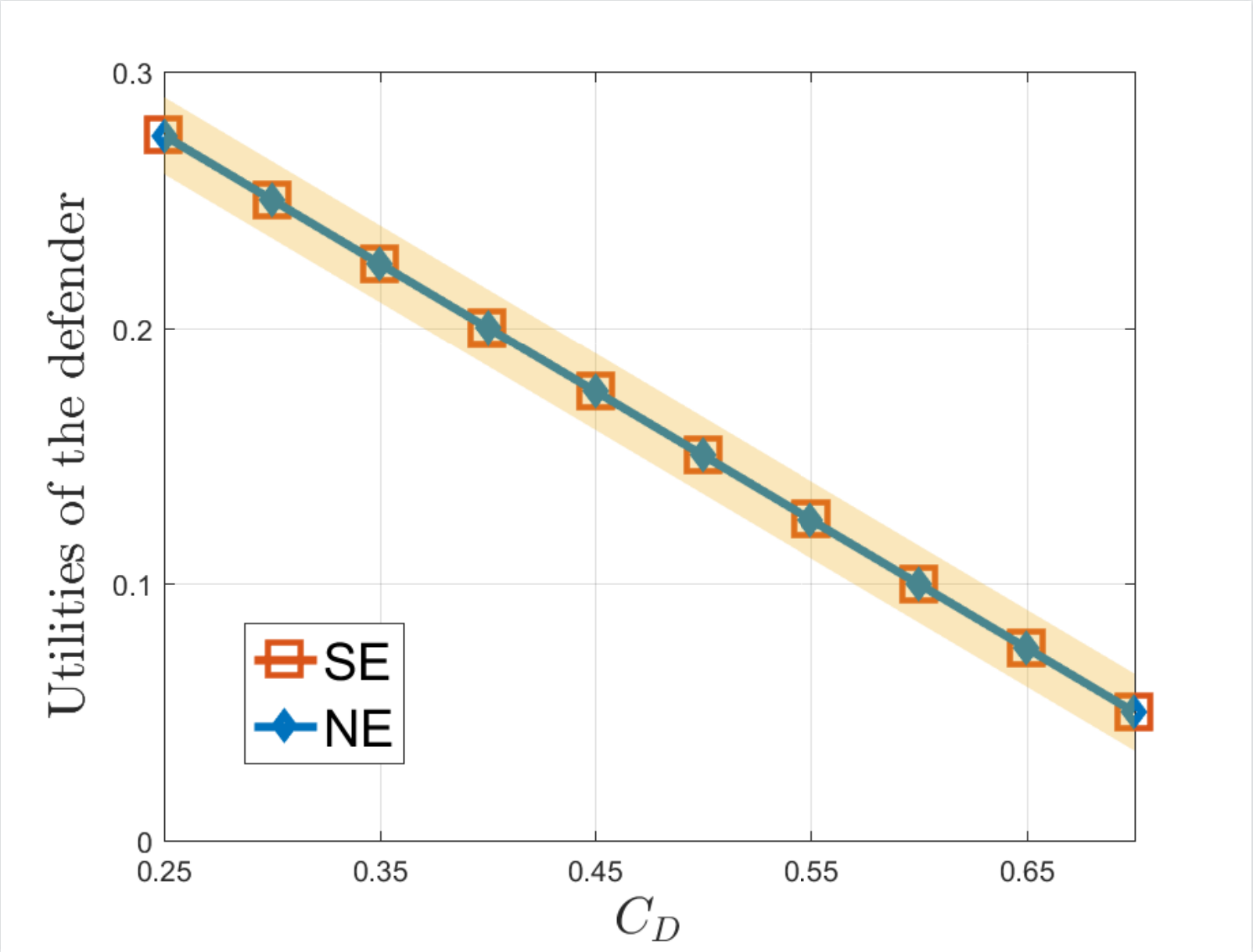}\\
 			\includegraphics[width=3.1cm]{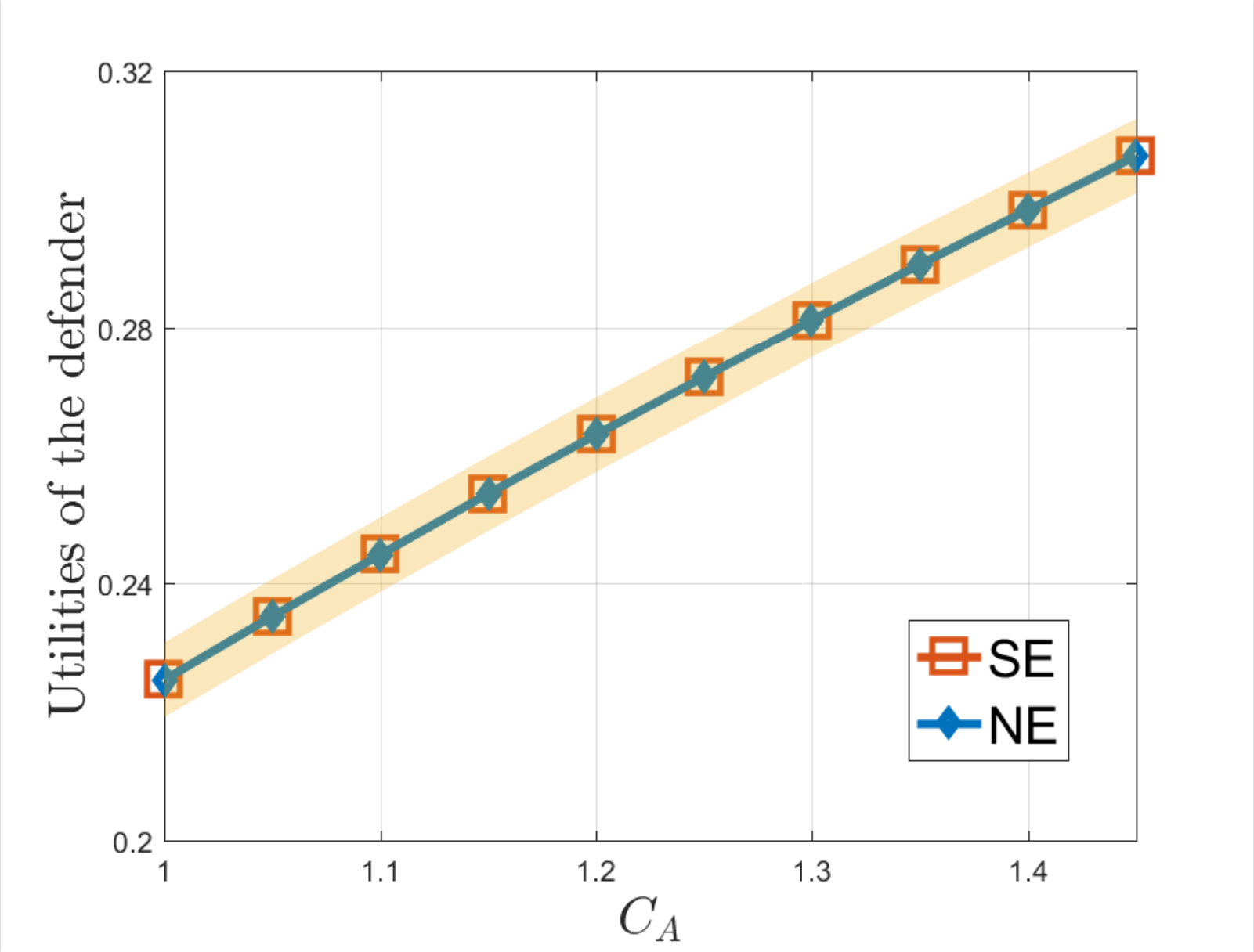}
 		\end{subfigure}%
 		%
 		\vspace{0.27cm}
 		\caption{Utilities of the defender}
 	\end{subfigure}
 	\centering
 	\caption{The relationship between SE and NE   with different environment settings.} 
 	\label{fig15}
 	\vspace{-0.65cm}
 \end{figure}
\subsection{Cooperative	secure transmission problems}


Consider a  secure transmission problem in a downlink heterogeneous network (HetNet), consisting of an MBS (player $\mathcal{X}$), a jamming SBS (player $\mathcal{Y}$), and a helping SBS (player $\mathcal{Z}$) \cite{wu2018secure,wu2016secure,wang2015pricing}. 
In the  leader-follower scheme,
the MBS  first determines  the amount of purchased  jamming power from the jamming SBS, and then the jamming SBS determines the associated service price, while the helping SBS finally determines the amount of the provided offloading service for the jamming SBS.
Denote $ x $ as the purchased jamming power  of MBS, $ y $  as the   price set by the jamming SBS for jamming service and offloading service, and $ z $ as the  amount of offloading service provided by the helping SBS. 
Denote $ R_{s} $ as the     secrecy rate, describing the difference between the achievable rate of the macrocell users and that of the eavesdropper. {It follows from reference  \cite{wu2018secure} that} the expression of $R_{s}$ is 
$ R_{M}- \log _{2}(1+\frac{P_M\left|g_{ Me}\right|^{2}/{N_{0}}}{1+ {x\left|g_{ je}\right|^{2}/{N_{0}}}+ \sigma_{k e}}),
$ 
where $ R_{M} $ is the
achievable rate at macrocell users,  $ P_{M} $ is the MBS's transmit powers,   $ \sigma_{k e} $ is the parameter related to transmitting powers of  the  unemployed
SBS,  $ \left|g_{ je}\right| $ and $ \left|g_{ Me}\right| $ are channel coefficients from the jamming SBS and the MBS, respectively, and   $ {N_{0}} $ is the variance of the additive white Gaussian noise.
%
%
%
%
%
%
{Referring to \cite{wu2018secure,wu2016secure,wang2015pricing}, the players' utility functions  are  described as}
\begin{equation*}\label{f15}
	\begin{aligned}
		\max_{x\in [x_{\operatorname{min}}, x_{\operatorname{max}}]}&  U_{\mathcal{X}}(x, y, z)=\lambda_{M} R_{s}- \left|g_{je}\right|^{2}xy,\\
		\max_{y\in [y_{\operatorname{min}}, y_{\operatorname{max}}]}&  U_{\mathcal{Y}}\left(x, y, z\right) \!=\!\left|g_{je}\right|^{2}xy-\theta x+\lambda_{j} \log \left(1+ z\right)-\!\tau y,\\
		\max_{z\in [z_{\operatorname{min}}, z_{\operatorname{max}}]}& U_{\mathcal{Z}}\left(x, y, z\right)=y^{2}\frac{z}{z+\alpha}-\omega z, \\
	\end{aligned}
\end{equation*}
where  $ \lambda_{M} $ denotes the unit profit for the secrecy rate,  $ \theta $  is the unit cost of the power consumption, $\tau$ is the economic incentive parameter, $ \lambda_{j} $ and $\alpha$  are  weighting factors, respectively, and $ \omega $ is the
unit cost. We denote 
%
$U_{\mathcal{X}}$ as 
the benefit  gained from the secrecy rate
and the payment  of employing the jamming SBS with $	B(x)= \lambda_{M} R_{s}$ and $f_{x}(y,z)x=- \left|g_{je}\right|^{2}xy$,
  $U_{\mathcal{Y}}$ as
   the reward of providing jamming service and  
  the diminishing benefit of offloading service with $f_{y1}(x,z)y=\left|g_{je}\right|^{2}xy-\!\tau y$ and $f_{y2}(x,z)=-\theta x+\lambda_{j} \log \left(1+ z\right)$,
and $U_{\mathcal{Z}}$    as the profit of offering offloading service.

 Accordingly, 
\vspace{-0.3cm}
 $$ 	
\begin{aligned}
	& T_{1}(x,y)=\! \left|g_{je}\right|^{2} x-\!\tau\!+\!\sqrt{\frac{\alpha}{\omega}}\frac{\lambda_{j}}{1\!-\!\alpha+\sqrt{\frac{\alpha}{\omega}}y},\quad\quad T_{2}(x)=\frac{\lambda_{M} \sigma_{M e}\left|g_{je}\right|^{2}}{\ln 2 N_{0}\left(\zeta_{x}^{2}+\sigma_{M e} \zeta_{x}\right)}\!-\!\frac{\left|g_{je}\right|^{2}\lambda_{j}}{(\tau\!-\!\left|g_{je}\right|^{2}x)^2},\vspace{-0.3cm}\\
	&T_{3}(x)=\left|g_{je}\right|^{2} x-\tau,\quad\quad \quad\quad \quad\quad\quad\quad \quad\quad \quad T_{4}(x)=\frac{\lambda_{M} \sigma_{M e}\left|g_{je}\right|^{2}}{\ln 2 N_{0}\left(\zeta_{x}^{2}+\sigma_{M e} \zeta_{x}\right)}-\left|g_{je}\right|^{2} y_{S E},\vspace{-0.3cm}
\end{aligned}
$$
where $\zeta_{x}=1+\left|g_{je}\right|^{2} x / N_{0}+\sigma_{k e}$ and $\sigma_{M e} =P_M\left|g_{ Me}\right|^{2}/{N_{0}}$.
\begin{figure}[htbp]
	\hspace{-1.5cm}
	\centering	
	\begin{subfigure}[t]{0.55\linewidth}
		\centering
		\includegraphics[width=5.2cm]{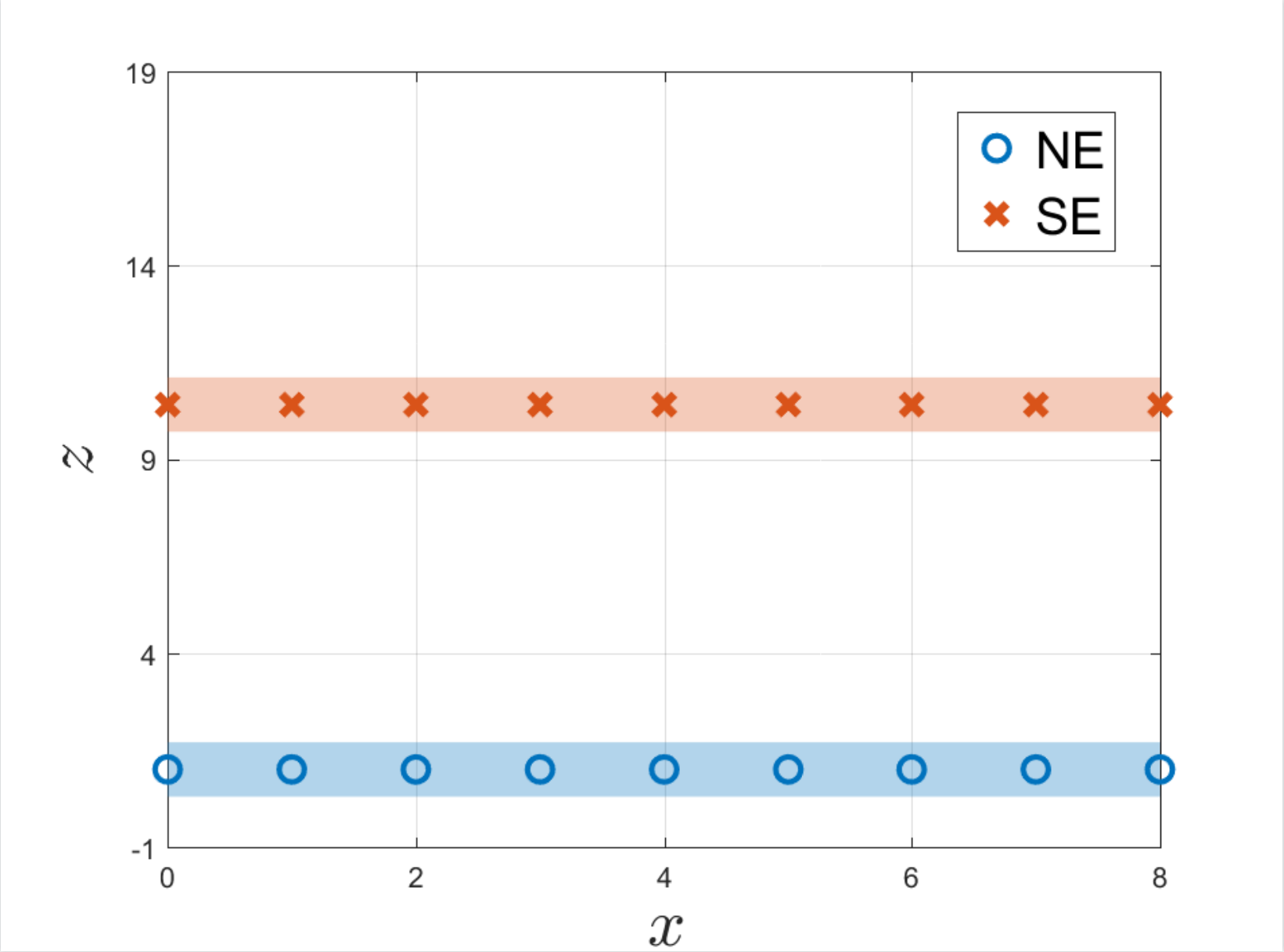}
		\caption{}
	\end{subfigure}%
	\hspace{-1.2cm}
	\begin{subfigure}[t]{0.55\linewidth}
		\centering
		\includegraphics[width=5.2cm]{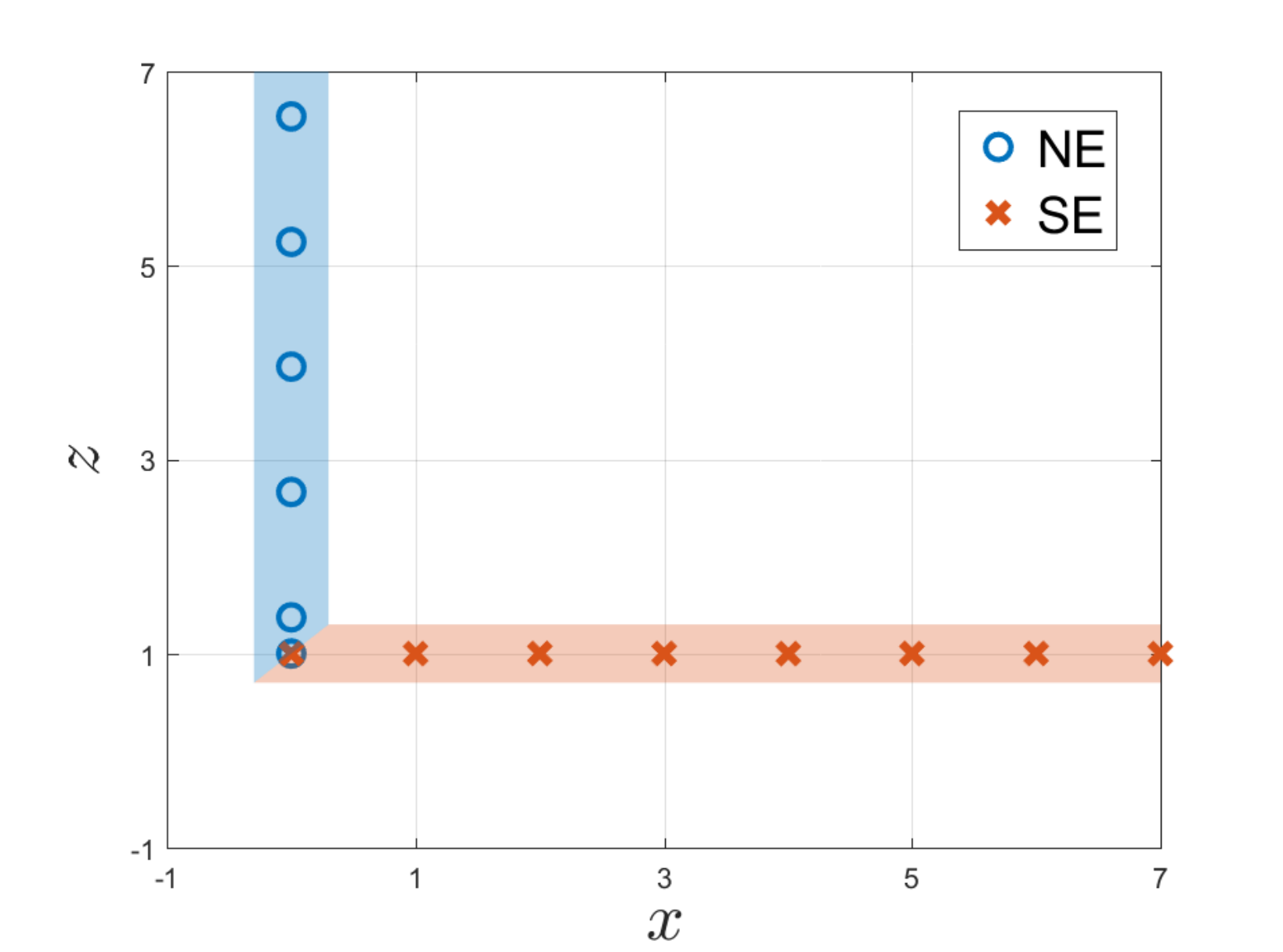}
		\caption{}
	\end{subfigure}%
	\centering
		\vspace{-0.2cm}
	\caption{To find   that at least one SE is an NE with different environment settings.} 
	\label{fig38}
	\vspace{-0.95cm}
\end{figure}
{In fact,  the information transmission in HetNets is more vulnerable to malicious eavesdropping attacks than  traditional single-tier networks, which makes  it challenging to obtain a satisfactory equilibrium for the MBS and SBSs  \cite{wu2016secure}.}
Consider the case that there exists an SE that is an NE. 
 From Theorem \ref{c2},  at least one SE is an NE of  $ \mathcal{G} $ if and only if there exists $  \delta(y_{SE}) $ and $  \delta(x_{SE}) $ such that $
y\geq     \frac{\lambda_{j}}{\tau-\left|g_{je}\right|^{2}x_{SE} }-\frac{\sqrt{\omega}(1-\alpha)}{\sqrt{\alpha}},$ for  $ y \in \delta(y_{SE})\cap \Omega_{\mathcal{Y}}
$ and  $  T_{4}(x)\cdot \operatorname{sgn}(x-x_{SE}) \leq 0 $ for $  x \in \delta(x_{SE})\cap \Omega_{\mathcal{X}}$.
Then we take two different environment parameter settings, and map the strategy spaces of all players on   the space $ \Omega_{\mathcal{X}}\times\Omega_{\mathcal{Z}} $ for clarification.
The red region represents the set of SE, while the blue region represents the set of NE.     In Fig. \ref{fig38}(a),
SE are always not NE by verifying the coincidence condition. Furthermore, in Fig. \ref{fig38}(b),  there is 
only one SE that meets an  NE. {It is usually hard  to reach this SE in secure transmission problems \cite{wu2018secure,wu2016secure,wang2015pricing}, since the probability
of finding such a singleton is zero.} However, by virtue of the derived
 condition in Theorem \ref{c2}, we can obtain this SE precisely and conveniently. In this way, the MBS  can commit to a   satisfactory SE strategy to enhance the security of the macrocell and guarantee user satisfaction, even when the channels may be interfered with external noise and the  SBSs’ observability may be lost.


On the other hand, we focus on the   closeness of SE and NE. Recalling that Theorem \ref{c7} gives an upper bound of  $H(\Xi_{SE},\Xi_{NE})$,
Fig. \ref{fig34}  reflects the variation trend of this bound under different environment settings.
 In Fig.   \ref{fig34},
the horizontal axis represents the value of $ \eta $ in Theorem \ref{c7}, while the vertical axis represents  the bound of $H(\Xi_{SE},\Xi_{NE})$,  expressed as  $\frac{(1+l)(\kappa_1+\kappa_2)}{\kappa_1\kappa_2}\eta$.
Set $P_M=15,30, 60, 120 (dBm)$ in Fig. \ref{fig34}(a) and set $\lambda_j=0.1,0.3,1,3$ in Fig. \ref{fig34}(b), which are involved in $ \kappa_1 $ and $ \kappa_2 $ in Theorem \ref{c7}, respectively.
These two environment parameters are important for secure transmission \cite{wu2016secure,wu2018secure}. As shown in Fig.  \ref{fig34}, the smaller value of $ \eta $ leads to the lower bound of $H(\Xi_{SE},\Xi_{NE})$.
 Also,  Fig. \ref{fig34}(a) and Fig. \ref{fig34}(b) show that  the performance gaps become small when $P_M$ and  $\lambda_j$ increase, as they serve as reciprocal terms, respectively, in $\frac{(1+l)(\kappa_1+\kappa_2)}{\kappa_1\kappa_2}\eta$.
{In a nutshell, different from \cite{wu2016secure},
	the  cooperation between the MBS and SBSs is further investigated from the equilibria relationship view. The  decline of the bound gap implies that although the MBS and the SBSs may not be in the same game scheme  due to the vulnerable transmission channels,  the brought conflict
	 can be ignored and the deviation between the SE strategy and  the NE strategy is tolerable. Hence,
	  the MBS and the SBSs can still achieve a win-win cooperation for the security enhancement. }
  \vspace{-0.7cm}
  \begin{figure}[htbp]
  	\hspace{-1.5cm}
  	\centering	
  	\begin{subfigure}[t]{0.55\linewidth}
  		\centering
  		\includegraphics[width=5.2cm]{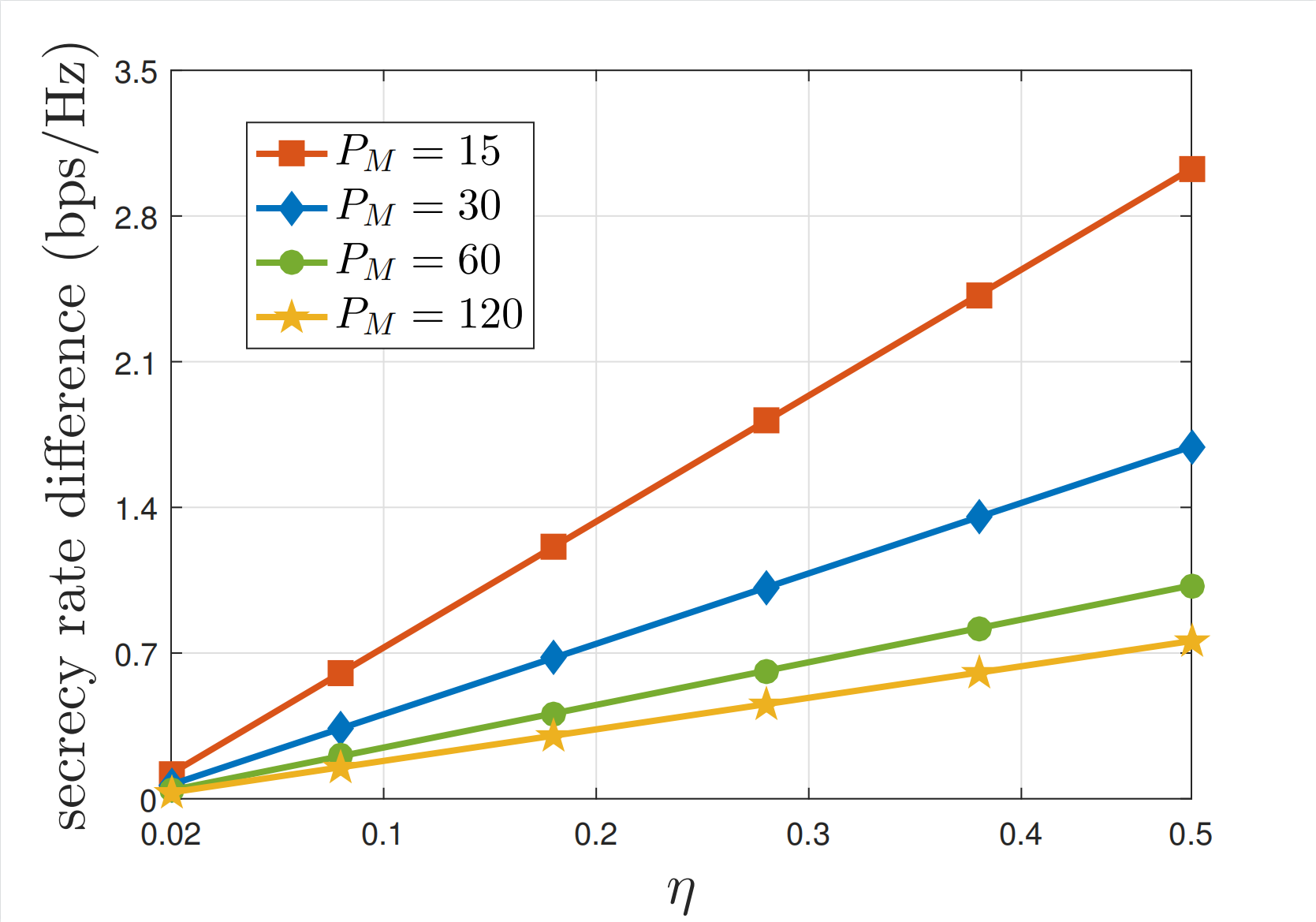}
  	\end{subfigure}%
  	\hspace{-1.2cm}
  	\begin{subfigure}[t]{0.55\linewidth}
  		\centering
  		\includegraphics[width=5.2cm]{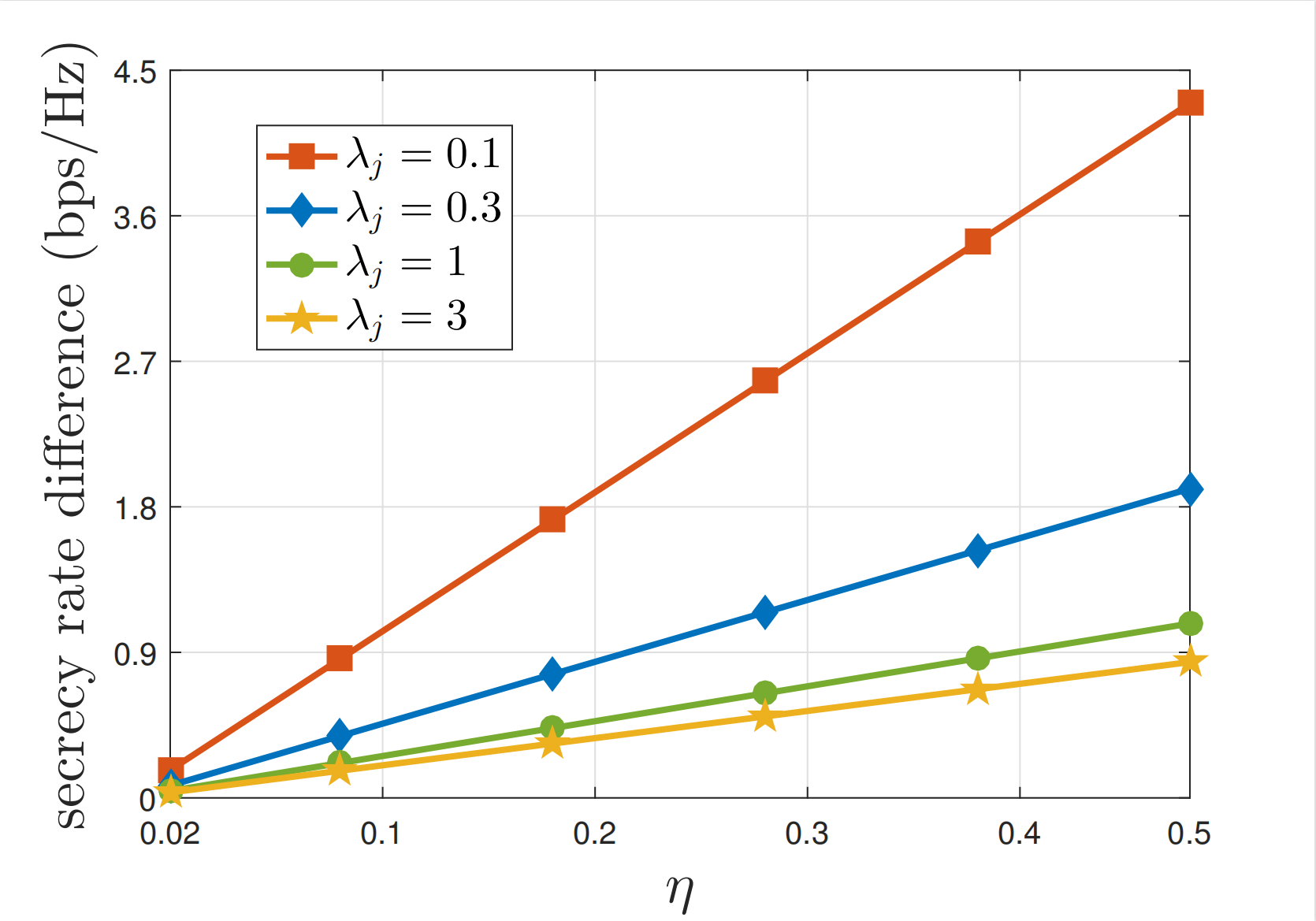}
  	\end{subfigure}%
  	\centering
  	\caption{Closeness of SE and NE in different environment settings.} 
  	\label{fig34}
  \end{figure}
  \vspace{-0.5cm}

\section{Conclusion}
In this paper, we have focused on a  three-player leader-follower security game and investigated the coincidence   between SE and NE.  We have provided a necessary and sufficient condition  such that each SE is an NE and presented the concise form when the SE is unique. {Besides, we  have provided another condition such that at least one SE is an NE.} Moreover, 
we have given an upper bound to measure their closeness once the coincidence condition fails.
Finally, we have shown the  
validity and  applicability of our results in several practical security cases. 


In the future, we may explore more deeply to extend the current research, including i) generalizing the model to $N$ players; ii)   quantitatively analyzing the influence of the uncertainty to the equilibrium; iii) investigating the equilibrium relationship for other game schemes.


%

\appendices
\section{
}\label{a11}

\textbf{Proof of Lemma \ref{l1}} Note that  $ \Omega_{\!\mathcal{X}}$, $ \Omega_{\!\mathcal{Y}}$, and $ \Omega_{\!\mathcal{Z}}$ are finite sets.    Since   $ BR_{y}\!(x) $ is a subset of $ \Omega_{\mathcal{Y}}$ for  $x\in\Omega_{\mathcal{X}}$ and $ BR_{z}\!(x,y) $ is a subset of $ \Omega_{\mathcal{Z}}$ for  $x\in\Omega_{\mathcal{X}}$ and $y\in\Omega_{\mathcal{Y}}$,  $ \mathcal{G} $ admits a Stackelberg strategy for player $\mathcal{X}$  \cite[Proposition 1]{bacsar1981equilibrium}. 
Therefore, there exists an SE of $ \mathcal{G} $.\hfill $\square$

\textbf{ Proof of  Lemma \ref{l2}}	Recalling (\ref{1a}a)-(\ref{1a}c), $ U_{\mathcal{X}}(x,y,z) $ is concave in $x$, $ U_{\mathcal{Y}}(x,y,z) $ is linear in $y$, and $ U_{\mathcal{Y}}(x,y,z) $ is concave in $z$, respectively. Together with the compactness and convexity of  $ \Omega_{\mathcal{X}}$, $ \Omega_{\mathcal{Y}}$ and $ \Omega_{\mathcal{Z}}$,  there 
exists an NE of $ \mathcal{G} $, referring to  \cite[Theorem 2.1]{carmona2012existence}.
\hfill $\square$



	\section{}\label{a16}
 \textbf{Proof of  Theorem \ref{t1}}  We first prove  the
	sufficiency.
	
	 Consider  $ y_{SE} $ and discuss coincidence condition (i)  in two cases:
	  $T_{3}(x_{SE})=0$ and $T_{3}(x_{SE})\!\neq\!0$.
	
	(1a) For the case that $ T_{3}(x_{SE})\!=\!\frac{\partial {U}_{\mathcal{Y}}(x_{SE},y,z_{SE})}{\partial y}\Big|_{y=y_{SE}}\!=\! 0 $, it is clear that
	$
	y_{SE} \!\in\! \underset{y\in \Omega_{\mathcal{Y}}}{\operatorname{argmax}}\,\! U_{\mathcal{Y}}(\!x_{SE}, $ $y, z_{SE})
	$ is player $ \mathcal{Y} $'s NE strategy due to the concavity of $U_{\mathcal{Y}}$ in $y$. 
	
	(1b) For the case that  $ T_{3}(x_{SE})\neq 0 $, consider that $ T_{3}(x_{SE})> 0 $ firstly. Together with  condition (i), we obtain
	 $ T_{1}(x_{SE},y) \geq0$ for $y\in \delta (y_{SE})\cap\Omega_{\mathcal{Y}}$.  
	Suppose that $y_{SE}\in \operatorname{rint}(\Omega_{\mathcal{Y}})$.
	 On the one hand, when $T_1(x_{SE},y_{SE})\neq 0$, we have $T_1(x_{SE},y_{SE})> 0$. Then due to the continuity of $T_1(x_{SE},y)=\frac{\partial \hat{U}_{\mathcal{Y}}(x_{SE},y)}{\partial y}$, 
	 there exists another point $y^{'}\in \delta_{+}(y_{SE})\cap \Omega_{\mathcal{Y}} $ such that 
	 $
	 \hat{U}_{\mathcal{Y}}(x_{SE}, y^{'})> \hat{U}_{\mathcal{Y}}\left(x_{SE}, y_{SE}\right). 
	 $
	 This contradicts the definition of $y_{SE}$.
	 On the other hand, 
	 when $T_1(x_{SE},y_{SE})= 0$,
	 there exists 
	 $\delta^{'}(y_{SE})$ such that 
	 $
	 \hat{U}_{\mathcal{Y}}(x_{SE}, y_{SE})> \hat{U}_{\mathcal{Y}}\left(x_{SE}, y\right) 
	 $  for $y\in \delta^{'}_{+}(y_{SE})\cap \Omega_{\mathcal{Y}} $, since $y_{SE}=BR_{y}(x_{SE})$ is a singleton. This
	  implies that
	 $T_1(x_{SE},y)<0$  for $y\in \delta^{'}_{+} (y_{SE})\cap\Omega_{\mathcal{Y}}$, which 
	  contradicts  condition (i).
	    Thus, 
	    $y_{SE}\notin \operatorname{rint}(\Omega_{\mathcal{Y}})$.
	      If $y_{SE}=y_{\operatorname{min}}$, then there exists  $\delta^{''}(y_{\operatorname{min}})$ such that $T_1(x_{SE},y)<0$ for $y \in \delta^{''}_{+}(y_{\operatorname{min}})\cap \Omega_{\mathcal{Y}}$. This also contradicts condition (i).
	  Therefore, 
	  $y_{SE}=y_{\operatorname{max}}$ is the only possible situation. Moreover, due to the concavity of $U_{\mathcal{Y}}$ in $y$, it follows from 
	  $ T_{3}(x_{SE})> 0 $ that $U_{\mathcal{Y}}\left(x_{SE}, y_{\operatorname{max}}, z_{SE}\right)\geq U_{\mathcal{Y}}\left(x_{SE}, y, z_{SE}\right) $ for $ y\in \Omega_{\mathcal{Y}}$. Thus, $y_{\operatorname{max}}$
 is  an NE strategy.
	 The analysis for the case that $ T_{3}(x_{SE})< 0 $ is similiar. Accordingly,  we obtain $y_{SE}=y_{\operatorname{min}}$, where $
	 y_{\operatorname{min}} \in \underset{y\in \Omega_{\mathcal{Y}}}{\operatorname{argmax}}\; U_{\mathcal{Y}}(x_{SE}, y, z_{SE})
	 $.
	 

	 Consider  $ x_{SE} $ and discuss condition (ii) in two cases: $T_{4}(x_{SE})=0$ and $T_{4}(x_{SE})\neq0$.
	 
	 (2a) For the case that $ T_{4}(x_{SE})=\frac{\partial {U}_{\mathcal{X}}(x,y_{SE},z_{SE})}{\partial x}\Big|_{x=x_{SE}}= 0 $, it is clear that
	 $
	 x_{SE} \in \underset{x\in \Omega_{\mathcal{X}}}{\operatorname{argmax}}\; U_{\mathcal{X}}(x,$ $y_{SE}, z_{SE}) 
	 $ is player $ \mathcal{X} $'s NE strategy  due to the concavity of $U_{\mathcal{X}}$ in $x$. 
	 
	 (2b) For the case that $ T_{4}(x_{SE})\neq 0 $,   with  
	 condition (ii),
	  we have $ T_{2}(x)\cdot T_{4}(x) >0 $ for $  x \in \delta(x_{SE})\cap \operatorname{rint}(\Omega_{\mathcal{X}}) $. If $x_{SE}\in \operatorname{rint}(\Omega_{\mathcal{X}})$, then  $T_{2}(x_{SE})=0$ due to the definition of $x_{SE}$, which contradicts condition (ii). 
	  Thus,  $ x_{SE}\notin \operatorname{rint}(\Omega_{\mathcal{X}}) $.
	   If  $x_{SE}=x_{\operatorname{max}}$, then there exists  $\delta^{'}(x_{\operatorname{max}})$ such that 
	 $T_{2}(x)\geq0$ for $x\in \delta^{'}(x_{\operatorname{max}})\cap \Omega_{\mathcal{X}}$. Thus,  $T_{2}(x)>0$ for $x\in \delta^{'}(x_{\operatorname{max}})\cap \operatorname{rint}(\Omega_{\mathcal{X}})$. Take $\delta^{''} (x_{\operatorname{max}})=\delta(x_{\operatorname{max}})\cap \delta^{'}(x_{\operatorname{max}})$.  Then we obtain
	 $  T_{4}(x) >0 $ for $  x \in \delta^{''}(x_{\operatorname{max}})\cap \operatorname{rint}(\Omega_{\mathcal{X}}) $. Moreover, due to the continuity and monotonicity of $T_{4}(x) $,  $  T_{4}(x) >0 $ for $  x \in  \operatorname{rint}(\Omega_{\mathcal{X}}) $. Therefore, 
	 $
	 x_{\operatorname{max}} \in \underset{x\in \Omega_{\mathcal{X}}}{\operatorname{argmax}}\; U_{\mathcal{X}}\left(x, y_{SE}, z_{SE}\right)
	 $, which indicates that $x_{\operatorname{max}}$ is an NE strategy.
	 The analysis for $x_{SE}=x_{\operatorname{min}}$ is similar, where
	 $
	 x_{\operatorname{min}} \in \underset{x\in \Omega_{\mathcal{X}}}{\operatorname{argmax}}\; U_{\mathcal{X}}\left(x, y_{SE}, z_{SE}\right)
	 $.
	 
	 Furthermore,  $ z_{SE}=  BR_{z}(x_{SE},y_{SE})$ of player $ \mathcal{Z} $  becomes  an NE strategy
	 when $ x_{SE}=x_{NE} $ and $y_{SE}=y_{NE}$.  Thus, when the concidence condition (i) and (ii) are satisfied, any SE is an NE. 
	 
	 	Next, we  prove  the
	 necessarity.	 
	 When $ (x_{SE},y_{SE},z_{SE}) $  is an NE, 
	 if $y_{SE}\in \operatorname{rint}(\Omega_{\mathcal{Y}})$, then  $ T_{3}(x_{SE})=0 $. This indicates that there exists $\delta(y_{SE})$  such that $T_{1}(x_{SE},y)\cdot T_{3}(x_{SE}) \geq0 $ for $  y \in \delta(y_{SE})\cap \Omega_{\mathcal{Y}}$. 
	 If $y_{SE}=y_{\operatorname{max}}$, then $ T_{3}(x_{SE})\geq0 $. Additionally, with the definition of SE,
	 there exists $\delta^{'}(y_{\operatorname{max}})$ such that $
	 \hat{U}_{\mathcal{Y}}(x_{SE}, y_{\operatorname{max}})\geq \hat{U}_{\mathcal{Y}}\left(x_{SE}, y\right) 
	 $ for   $y \in \delta^{'}(y_{\operatorname{max}})\cap \Omega_{\mathcal{Y}}$, which yields
	 $T_{1}(x_{SE},y)=\frac{\partial \hat{U}_{\mathcal{Y}}(x_{SE},y)}{\partial y}\geq0 $ for $  y \in \delta^{'}(y_{\operatorname{max}})\cap \Omega_{\mathcal{Y}}$.
	 Thus, 
	  $T_{1}(x_{SE},y)\cdot T_{3}(x_{SE}) \geq0, $ for $  y \in \delta^{'}(y_{\operatorname{max}})\cap \Omega_{\mathcal{Y}}$.  The analysis for the case that $y_{SE}=y_{\operatorname{min}}$ is similar. 
	 
	 On the other hand, if $x_{SE}\in \operatorname{rint}(\Omega_{\mathcal{X}})$, then $ T_{4}(x_{SE})=0 $. Moreover, when  $x_{SE}=x_{\operatorname{max}}$, if $T_{4}(x_{\operatorname{max}})=0$, then condition (ii) is satisfied. If not,
	 then there exists $\delta^{'}(x_{\operatorname{max}})$ such that $T_{4}(x)>0 $ for $  x \in \delta^{'}(x_{\operatorname{max}})\cap \operatorname{rint}(\Omega_{\mathcal{X}})$. Moreover, recalling the definition of SE,
	  there exists  $\delta^{''}(x_{\operatorname{max}})$ such that
	  $
	  \hat{U}_{\mathcal{X}}(x_{\operatorname{max}})> \hat{U}_{\mathcal{X}}\left(x\right) 
	  $ 
	for $  x \in \delta^{''}(x_{\operatorname{max}})\cap \operatorname{rint}(\Omega_{\mathcal{X}})$, which yields $T_{2}(x)=\frac{\partial \hat{U}_{\mathcal{X}}(x)}{\partial x}>0 $  for $  x \in \delta^{''}(x_{\operatorname{max}})\cap \operatorname{rint}(\Omega_{\mathcal{X}})$. 
	 Thus, by taking $\delta(x_{\operatorname{max}})=\delta^{'}(x_{\operatorname{max}})\cap\delta^{''}(x_{\operatorname{max}})$, we have $T_{2}(x)\cdot T_{4}(x) >0 $ for $  x \in \delta(x_{\operatorname{max}})\cap \operatorname{rint}(\Omega_{\mathcal{X}}) $. The analysis for $x_{SE}=x_{\operatorname{min}}$ is similar. \hfill $\square$
%
%
%
%
%
%
%
%
%

 \section{}\label{a18}
 \textbf{Proof of Corollary  \ref{c1}}
 Notice that the coincidence condition for $y_{SE} $ in Corollary \ref{c1} is the same as that in Theorem \ref{t1}, so we omit it and focus on $x_{SE}$. 
 
Consider the
 sufficiency firstly. If $  x_{SE} \in \operatorname{rint}(\Omega_{\mathcal{X}}) $, then  
  $ T_{2}(x_{SE}) 
 =\frac{\partial \hat{U}_{\mathcal{X}}(x)}{\partial x}\Big|_{x=x_{SE}}
 =0 $. Moreover, due to the uniqueness of $x_{SE}$, 
  there exists $\delta^{'}(x_{SE})$ such that $T_{2}(x)>0$ for $x\in  \delta^{'}_{-} (x_{SE})\cap\Omega_{\mathcal{X}}$ and $T_{2}(x)<0$ for $x\in  \delta^{'}_{+} (x_{SE})\cap\Omega_{\mathcal{X}}$. 
  Together with condition (ii), by taking $\delta^{''}(x_{SE})=\delta(x_{SE})\cap\delta^{'}(x_{SE})$, we obtain 
 $T_{4}(x)\geq0$ for $x\in  \delta^{''}_{-} (x_{SE})\cap\Omega_{\mathcal{X}}$, $T_{4}(x)\leq0$ for $x\in  \delta^{''}_{+} (x_{SE})\cap\Omega_{\mathcal{X}}$ and  $T_{4}(x_{SE})= 0 $.
 Based on the concavity and continuity of ${U}_{\mathcal{X}}$, we further have 
 $T_{4}(x)\geq0$ for $x\in  [x_{\operatorname{min}}, x_{SE})$ and
 $T_{4}(x)\leq0$ for $x\in  (x_{SE},x_{\operatorname{max}}]$. Thus, it is clear that $x_{SE}$  is player $ \mathcal{X} $'s NE strategy. 
 If  $x_{SE}=x_{\operatorname{max}}$, then $T_{2}(x_{\operatorname{max}})\geq0$, and
 there exists 
 $\delta(x_{\operatorname{max}})$ such that $T_{2}(x)>0 $ for $  x \in \delta(x_{\operatorname{max}})\cap \operatorname{rint}(\Omega_{\mathcal{X}})$. With condition (ii), denote $\delta^{''}(x_{\operatorname{max}})=\delta(x_{\operatorname{max}})\cap\delta^{'}(x_{\operatorname{max}})$. Then  $T_{4}(x)\geq0$ for $x\in  \delta^{''} (x_{\operatorname{max}})\cap\Omega_{\mathcal{X}}$. Obviously, $T_{4}(x)\leq0$ for $x\in\Omega_{\mathcal{X}}$, which indicates that $x_{\operatorname{max}}$  is player $ \mathcal{X} $'s NE strategy.
 Besides, the analysis for
  $x_{SE}=x_{\operatorname{min}} $ is similar. 
 
 Next, consider the
 necessarity.  If 
 $x_{SE}$ is an NE strategy, then  $
 x_{SE}\! \in\! \underset{x \in \Omega_{\mathcal{X}}}{\operatorname{argmax}}\, U_{\mathcal{X}}(x,$ $ y_{SE}, z_{SE})$.
 If $x_{SE}\in \operatorname{rint}(\Omega_{\mathcal{X}})$, then   $ T_{4}(x_{SE})=0 $, $T_{4}(x)\geq0$ for $x\in  [x_{\operatorname{min}}, x_{SE})$ and
 $T_{4}(x)\leq0$ for $x\in  (x_{SE},x_{\operatorname{max}}]$ due to the monotonicity of $T_{4}$.
 Also, since the SE is unique,  it is clear that $ T_{2}(x_{SE})=0 $, and
 there exists $\delta(x_{SE})$ such that $T_{2}(x)>0 $ for $  x \in \delta_{-}(x_{SE})\cap \Omega_{\mathcal{X}}$, $T_{2}(x)<0 $ for $  x \in \delta_{+}(x_{SE})\cap \Omega_{\mathcal{X}}$. Thus, 
  $T_{2}(x)\cdot T_{4}(x) \geq0 $ for $  x \in \delta(x_{SE})\cap \Omega_{\mathcal{X}}$.  If  $x_{SE}=x_{\operatorname{max}}$, then $T_{4}(x)\geq0 $ for $  x \in  \Omega_{\mathcal{X}}$. Moreover, $T_{2}(x_{\operatorname{max}})\geq0$, and
  there exists 
 $\delta(x_{\operatorname{max}})$ such that $T_{2}(x)>0 $ for $  x \in \delta(x_{\operatorname{max}})\cap \operatorname{rint}(\Omega_{\mathcal{X}})$. Therefore,  $T_{2}(x)\cdot T_{4}(x) \geq0 $ for $  x \in \delta(x_{\operatorname{max}})\cap \Omega_{\mathcal{X}}$. Moreover, the analysis for  $x_{SE}=x_{\operatorname{min}}$ is similar. \hfill $\square$

\section{}\label{a19}
\textbf{Proof of Theorem \ref{c2}}
 The coincidence condition for $y_{SE} $ in Theorem \ref{c2} is the same as that  in Theorem \ref{t1}, so we omit it and focus on the analysis of $x_{SE}$. 
 
  Consider the
 sufficiency firstly.  If $  x_{SE} \in \operatorname{rint}(\Omega_{\mathcal{X}}) $, then $ \operatorname{sgn}(x_{SE}-x_{SE}) =0$,  $ \operatorname{sgn}(x-x_{SE}) =-1$ for  $x\in \delta_{-} (x_{SE})\cap\Omega_{\mathcal{X}}$, and  $ \operatorname{sgn}(x-x_{SE}) =1$ for  $x\in \delta_{+} (x_{SE})\cap\Omega_{\mathcal{X}}$.  Together with condition (ii) of Theorem \ref{c2}, it derives that $T_4(x)\geq0$ for $x\in \delta_{-} (x_{SE})\cap\Omega_{\mathcal{X}}$ and $T_4(x)\leq0$ for $x\in \delta_{+} (x_{SE})\cap\Omega_{\mathcal{X}}$. In this way,  $T_4(x_{SE})=0$ due to the continuity of $T_4$, which implies that $x_{SE}$  is player $ \mathcal{X} $'s NE strategy.
  If $x_{SE}=x_{\operatorname{max}}  $, then $ \operatorname{sgn}(x_{\operatorname{max}}-x_{\operatorname{max}}) =0$,  $ \operatorname{sgn}(x-x_{\operatorname{max}}) =-1$ for  $x\in \delta_{-} (x_{\operatorname{max}})\cap\Omega_{\mathcal{X}}$. Similarly, we have $T_4(x_{\operatorname{max}})=0$, which implies that $x_{\operatorname{max}}$  is player $ \mathcal{X} $'s NE strategy. Also, the analysis for $x_{SE}=x_{\operatorname{min}}$ is similar. 
 
  Next, consider the
 necessarity. When there exists an SE which is an NE, if  $x_{SE}\in \operatorname{rint}(\Omega_{\mathcal{X}})$, then  $T_4(x_{SE})=0$, and there exists $\delta (x_{SE})$ such that $T_4(x)\geq0$ for $x\in \delta_{-} (x_{SE})\cap\Omega_{\mathcal{X}}$, 
 $T_4(x)\leq0$ for $x\in \delta_{+} (x_{SE})\cap\Omega_{\mathcal{X}}$. Moreover, we have  $ \operatorname{sgn}(x_{SE}-x_{SE}) =0$,  $ \operatorname{sgn}(x-x_{SE}) =-1$ for  $x\in \delta_{-} (x_{SE})\cap\Omega_{\mathcal{X}}$, and  $ \operatorname{sgn}(x-x_{SE}) =1$ for  $x\in \delta_{+} (x_{SE})\cap\Omega_{\mathcal{X}}$.  Thus,  $  T_{4}(x)\cdot \operatorname{sgn}(x-x_{SE}) \leq 0 $ for $  x \in \delta(x_{SE})\cap \Omega_{\mathcal{X}}$. If  $x_{SE}=x_{\operatorname{max}}$, then there exists $\delta (x_{\operatorname{max}})$ such that $T_4(x)\geq0$ for $x\in \delta (x_{\operatorname{max}})\cap\Omega_{\mathcal{X}}$. Also,   $ \operatorname{sgn}(x_{\operatorname{max}}-x_{\operatorname{max}}) =0$,  $ \operatorname{sgn}(x-x_{\operatorname{max}}) =-1$ for  $x\in \delta (x_{\operatorname{max}})\cap\Omega_{\mathcal{X}}$. Hence, condition (ii) is also satisfied. Moreover, the analysis for $x_{SE}=x_{\operatorname{min}}$ is similar. \hfill $\square$
\section{}\label{a21}
\textbf{Proof of Theorem \ref{c7}}
Recall  that $BR_{z}\!(x_{SE},y_{SE})\! =\!z_{SE}$ and $BR_{z}\!(x_{NE},y_{NE}) \!=\!z_{NE}$.
Because $ \Omega_{\mathcal{Y}} $ and $ \Omega_{\mathcal{X}} $ are compact  with
$ BR_{z}(x,y) \!\in\! \mathcal{C}^{1} $ in $x$ and $y$, there exists a constant $l>0$ such that  
\vspace{-0.4cm}
$$
	\|BR_{z}(x_{SE},y_{SE})- BR_{z}(x_{NE},y_{NE})\|
	 \leq l (\|x_{SE}-x_{NE} \| + \|y_{SE}-y_{NE} \|).
	\vspace{-0.4cm}
$$
Since $\hat{U}_{\mathcal{Y}}$ is $\kappa_1$-strongly concave in $y$,  
$
\kappa_1 \|y_{SE}-y_{NE} \|\leq \|T_1(x_{SE},y_{SE})-T_1(x_{SE},y_{NE}) \|.
$
Also, since $\hat{U}_{\mathcal{X}}$ is $\kappa_2$-strongly concave in $x$, 
$
\kappa_2 \|x_{SE}-x_{NE} \|\leq \|T_2(x_{SE})-T_2(x_{NE}) \|.
$ Let $p^{*}=(x_{SE},y_{SE},z_{SE})$ be an SE    and  $q^{*}=(x_{SE},y_{SE},z_{SE})$ be an NE. 
Then  we have
$
\|p^{*}-q^{*}\|
\leq(1+l)  (\|x_{SE}-x_{NE} \|+\|y_{SE}-y_{NE} \|).
$
Following  the defintion of  $H(\Pi_{x_{SE}},\Pi_{x_{NE}})$, we get 
\vspace{-0.5cm}
\begin{align*}
	H(\Pi_{x_{SE}},\Pi_{x_{NE}})&= \max\{\sup_{T_2(x_{SE})\in \Pi_{x_{SE}}}\inf_{T_2(x_{NE})\in \Pi_{x_{NE}}}\|T_2(x_{SE})-T_2(x_{NE})\|,\\\vspace{-0.5cm}
		&	\quad\quad\quad\;\,\sup\limits_{T_2(x_{NE})\in \Pi_{x_{NE}}} \inf \limits_{T_2(x_{SE})\in \Pi_{x_{SE}}} \|T_2(x_{SE})-T_2(x_{NE})\| \}\\\vspace{-0.5cm}
			&\geq  \max\{\sup_{T_2(x_{SE})\in \Pi_{x_{SE}}}\inf_{T_2(x_{NE})\in \Pi_{x_{NE}}}\kappa_2\|x_{SE}-x_{NE}\|,\\\vspace{-0.5cm}
	&	\quad\quad\quad\;\,\sup\limits_{T_2(x_{NE})\in \Pi_{x_{NE}}} \inf \limits_{T_2(x_{SE})\in \Pi_{x_{SE}}} \kappa_2\|x_{SE}-x_{NE}\| \}.\vspace{-0.5cm}
\end{align*}	   
Similarly, $	H(\Pi_{y_{SE}},\Pi_{y_{NE}})
\geq  \max\{\sup\limits_{T_1(x_{SE},y_{SE})\in \Pi_{y_{SE}}}\inf\limits_{T_1(x_{SE},y_{NE})\in \Pi_{y_{NE}}}\kappa_1\|y_{SE}-y_{NE}\|,$ 
$
\sup\limits_{T_1(x_{SE},y_{SE})\in \Pi_{y_{NE}}}$ $\inf \limits_{T_1(x_{SE},y_{NE})\in \Pi_{y_{SE}}} \kappa_1\|y_{SE}-y_{NE}\| \}$.
\vspace{0.2cm}

Therefore,    
\vspace{-0.5cm}
\begin{align*}
	H(\Xi_{SE},\Xi_{NE})&= \max\{\sup_{p^{*} \in \Xi_{SE} }\inf_{q^{*} \in \Xi_{NE}}\|p^{*}-q^{*}\|,\sup\limits_{q^{*} \in \Xi_{NE}} \inf \limits_{p^{*} \in \Xi_{SE}} \|p^{*}-q^{*}\| \}\vspace{-0.8cm}\\
		&	\leq {(1+l)}/{\kappa_1}\max\{\sup_{T_1(x_{SE},y_{SE}) \in \Pi_{y_{SE} } }\inf_{T_1(x_{NE},y_{SE})  \in \Pi_{y_{NE} }} \|T_1(x_{SE},y_{SE})-T_1(x_{SE},y_{NE}) \|,\\\vspace{-0.8cm}
			& \quad\quad\quad\quad\quad\quad\,\,\sup\limits_{T_1(x_{NE},y_{SE}) \in \Pi_{y_{NE} }} \inf \limits_{T_1(x_{SE},y_{SE}) \in \Pi_{y_{SE} }}  \|T_1(x_{SE},y_{SE})-T_1(x_{SE},y_{NE}) \| \}\\\vspace{-0.8cm}		
						&+
			{(1+l)}/{\kappa_2} \max\{\sup_{T_2(x_{SE})  \in \Pi_{x_{SE} } }\inf_{T_2(x_{NE})   \in \Pi_{x_{NE} }} \|T_2(x_{SE})-T_2(x_{NE}) \|,\\\vspace{-0.8cm}	
				&  \quad\quad\quad\quad\quad\quad\,\,\sup\limits_{T_2(x_{NE})   \in \Pi_{x_{NE} }} \inf \limits_{T_2(x_{SE})  \in \Pi_{x_{SE} }}  \|T_2(x_{SE})-T_2(x_{NE}) \| \}\\ \vspace{-0.8cm}
					& ={(1+l)}/{\kappa_1}\;	H(\Pi_{y_{SE}},\Pi_{y_{NE}})+{(1+l)}/{\kappa_2}\;	H(\Pi_{x_{SE}},\Pi_{x_{NE}})	\\   \vspace{-0.8cm}
	&\leq \frac{(1+l)(\kappa_1+\kappa_2)}{\kappa_1\kappa_2}\eta,\vspace{-0.7cm}
\end{align*}	   
which yields the conclusion.  \hfill$\square$



\ifCLASSOPTIONcaptionsoff
  \newpage
\fi

	\bibliographystyle{IEEEtran}
\bibliography{hyper}

\end{document}